\documentclass[preprint,1p]{elsarticle}

\usepackage{graphicx}% Include figure files
\usepackage{amsmath}
\usepackage{comment}% Include the comment package
\usepackage{appendix}
\biboptions{sort&compress}

\begin{document}

\author[1]{Marko Korhonen\corref{cor1}}
\ead{marko.korhonen@aalto.fi}
\author[1]{Alpo Laitinen}
\ead{alpo.laitinen@aalto.fi}
\author[1]{Gizem Ersavas Isitman}
\ead{gizem.ersavasisitman@aalto.fi}
\author[2]{Jose L. Jimenez}
\ead{jose.jimenez@Colorado.EDU}
\author[1]{Ville Vuorinen}
\ead{ville.vuorinen@aalto.fi}

\cortext[cor1]{Corresponding author}

\address[1]{Department of Mechanical Engineering, Aalto University, P.O. Box 14100 FI-00076 AALTO, Finland}
\address[2]{Department of Chemistry and CIRES, University of Colorado, 216 UCB, Boulder, CO 80309-0216, USA}

\title{A GPU-accelerated computational fluid dynamics solver for assessing shear-driven indoor airflow and virus transmission by scale-resolved simulations}

\date{\today}

\begin{abstract}
We explore the applicability of MATLAB for 3D computational fluid dynamics (CFD) of shear-driven indoor airflows. A new scale-resolving, large-eddy simulation (LES) solver titled DNSLABIB is proposed for MATLAB utilizing graphics processing units (GPUs). In DNSLABIB, the finite difference method is applied for the convection and diffusion terms while a Poisson equation solver based on the fast Fourier transform (FFT) is employed for the pressure. The immersed boundary method (IBM) for Cartesian grids is proposed to model solid walls and objects, doorways, and air ducts by binary masking of the solid/fluid domains. The solver is validated in two canonical reference cases. Then, we demonstrate the validity of DNSLABIB in a room geometry by comparing the results against another CFD software (OpenFOAM). Next, we demonstrate the solver performance in several isothermal indoor ventilation configurations and the implications of the results are discussed in the context of airborne transmission of COVID-19. The novel numerical findings using the new CFD solver are as follows. First, a linear scaling of DNSLABIB is demonstrated and a speed-up by a factor of 3-4 is also demonstrated in comparison to similar OpenFOAM simulations. Second, ventilation in three different indoor geometries are studied at both low (0.1m/s) and high (1m/s) airflow rates corresponding to $Re=5000$ and $Re=50000$. An analysis of the indoor CO$_2$ concentration is carried out as the room is emptied from stale, high CO$_2$ content air. We estimate the air changes per hour (ACH) values for three different room geometries and show that the numerical estimates from 3D CFD simulations may differ by 80--150 \% ($Re=50000$) and 75--140 \% ($Re=5000$) from the theoretical ACH value based on the perfect mixing assumption. Third, the analysis of the CO$_2$ probability distributions (PDFs) indicates a relatively non-uniform distribution of fresh air indoors. Fourth, utilizing a time-dependent Wells-Riley analysis, an example is provided on the growth of the cumulative infection risk which is shown to reduce rapidly after the ventilation is started. The average infection risk is shown to reduce by a factor of 2 for lower ventilation rates (ACH=3.4-6.3) and 10 for the the higher ventilation rates (ACH=37-64). Finally, we utilize the new solver to comment on respiratory particle transport indoors. The primary contribution of the paper is to provide an efficient, GPU compatible CFD solver environment enabling scale-resolved simulations (LES/DNS) of airflow in large indoor geometries on a desktop computer. The demonstrated efficacy of MATLAB for GPU computing indicates a high potential of DNSLABIB for various future developments on airflow prediction.  
\end{abstract}


\maketitle

%%%MAIN TEXT%%%%
\section{Introduction}
The COVID-19 pandemic has set an unprecedented demand for multidisciplinary research to comprehend the transmission mechanisms of the SARS-CoV-2 virus~\cite{wang2021airborne}.
At the onset of the pandemic, the virus was initially assumed to transmit predominantly via larger droplets and fomites present on surfaces~\cite{wang2021airborne}. However, since the early 2020, consistent and mounting evidence on the airborne transmission of the SARS-CoV-2 virus has accumulated ~\cite{tellier2022covid,auvinen2022high,anderson2020consideration,tang2020aerosol,jayaweera2020transmission,mittal2020flow,fears2020persistence,van2020aerosol,zhang2020identifying,wilson2020airborne,godri2020covid,li2021probable,henriques2021modelling}. Within the aerosol physics community, the suspension time of airborne particles in air has been well-established for a century. Therefore, during the pandemic, the fluid physics research community has further revisited various factors affecting particle transport in the air. These include the impact of particle size on their ability to remain airborne, the effect of relative humidity on particle shrinkage as well as particle transport over large distances in turbulent indoor airflow~\cite{eames2022spread}. Indeed, early scientific contributions on the airborne transmission of the virus were provided by the physics-based computational fluid dynamics (CFD) simulations, which addressed the airborne transport of small particles in different indoor settings. For instance, Ascione~{\it{et al.}} conducted a comprehensive study on the effects of various HVAC retrofitting alternatives in a university faculty which included CFD simulations on the ventilation configurations~\cite{ascione2021design}. Zhang~{\it{et al.}} performed CFD analysis of humidity and temperature distributions and ventilation performance in an indoor space using Ansys FLUENT~\cite{zhang2021simulation}. Abuhegazy {\it{et al.}} provided a CFD simulation of a classroom in FLUENT and a detailed discussion on how windows, glass barriers as well as aerosol size and source might effect the particle trajectories~\cite{abuhegazy2020numerical}. Other CFD studies have considered the impact of ventilation on the distribution of aerosols from coughing using a commercial software~\cite{borro2021role} and particle trajectories in OpenFOAM~\cite{liu2020modeling} as well as utilizing far-UVC lightning as a virus inactivator~\cite{buchan2020predicting}. Furthermore, various elements impacting the spread of airborne particles, such as ventilation, air filters and masks, have been considered in assorted CFD publications~\cite{vuorinen2020modelling,ren2021numerical,dbouk2021airborne,ho2021modeling,ho2021modelling,li2020investigating,khosronejad2020fluid}. At present, the aerosol inhalation route is broadly acknowledged to be one of the key mechanisms, possibly the main mechanism, of SARS-CoV-2 transmission ~\cite{who2021,wang2021airborne,tellier2022covid}. 
 
From the modeling perspective, Reynolds-averaged Navier-Stokes (RANS) modeling has been favored over scale-resolved simulations in indoor airflow simulations in the past despite the reduced accuracy and the evidence to its inability to capture essential turbulent phenomena in such flows~\cite{nielsen2015fifty,blocken2018over}.
While large-eddy simulation (LES) and direct numerical simulation (DNS) can certainly mitigate these problems, these methods naturally imply more computational effort due to increased mesh sizes and level of complexity. In order to promote such scale-resolving approaches in realistic indoor airflow modeling, efficient computational approaches are therefore required. 

In this context, the advances in GPU based computing may prove increasingly beneficial for these large simulations, since their architecture is well suited for performing parallel computations on large numerical systems~\cite{pratx2011gpu} and their value specifically for CFD has also been established~\cite{niemeyer2014recent,liu2004real}. In addition to incompressible Navier-Stokes solvers~\cite{scheidegger2005practical,shinn2009implementation,thibault2009cuda,brandvik2011turbo}, successful GPU implementations have been produced for multiphase flows~\cite{griebel2010multi,zaspel2013solving,kelly2014numerical}, direct numerical simulations~\cite{shinn2010direct,salvadore2013gpu,khajeh2013direct} and reactive flows~\cite{shi2012accelerating,spafford2009accelerating}, for instance.  Of the most recent work, GPU enabled CFD simulations based on the concept of artificial compressibility method has been demonstrated in {\it{PyFr}}~\cite{witherden2014pyfr,vermeire2017utility,loppi2018high}, which has been since augmented with optimal Runge-Kutta schemes~\cite{vermeire2019optimal} and locally adaptive pseudo-time stepping~\cite{loppi2019locally} for increased performance. Additionally, in~\cite{vuorinen2016dnslab}, a modified Chorin-Temam projection method was implemented with spectral methods and utilized in solving the Navier-Stokes equations in a periodic flow geometry on a CPU in 2D. During the pandemic, the code (DNSLAB) by Vuorinen et al.~\cite{vuorinen2016dnslab} was extended by the authors to 3D and rendered compatible with GPUs for periodic flows without walls. The multidisciplinary research consortium work by Vuorinen {\it{et al.}} in the spring of 2020 was among the first systematic CFD assessments of COVID-19 airborne transmission~\cite{vuorinen2020modelling}. These investigations, using several open-source CFD codes, implied that the DNSLAB runs on a GPU clearly outperformed OpenFOAM simulations on a supercomputer in terms of computational time for simple problem types, {\it{i.e.}} fully periodic flows. 

While the computational capabilities of GPUs for CFD have certainly been recognized by many, the inherent power of these devices can be offset by the steeply increasing requirement for technical expertise as the efficient implementation of the system of equations generally involves a suitable API, such as CUDA~\cite{nickolls2008scalable}. Therefore, a more simplified software environment for these GPU implementations, such as MATLAB, would be preferable for the majority of users. Indeed, MATLAB has been endorsed in many other fields of scientific computing, including neuroscience~\cite{peyk2011electromagnetoencephalography,goodman2009brian}, modeling of electrical circuits and systems~\cite{daowd2011passive,mohanty2014matlab,alegre2017modelling} and control and communication systems~\cite{gu2005robust,wang2009model,proakis2012contemporary,cho2010mimo}. However, similar adoption of the software in the CFD community has not been materialized and as a result, its increasing potential as an accessible computational tool may therefore be neglected. In our view, a streamlined LES simulation software with the capacity to solve very large systems rapidly in MATLAB is therefore warranted.

Hence, in an effort to bridge the research gap, we present a GPU compatible CFD solver for shear-driven airflow problems in simplified geometries. In our software, ease of use and performance are emphasized to allow scale-resolved turbulent flow simulations (similar to LES and DNS) in typical indoor environments involving ventilation airflow, for instance.
The main objectives of the paper are as follows.
First, to explore the possibility of performing incompressible 3D scale-resolved flow simulations on a GPU in MATLAB. 
Second, to implement and validate a simplified immersed boundary (IB) method in MATLAB in order to explore indoor settings with solid obstacles, walls, tables and furniture. 
Third, to employ the new solver to characterize indoor airflow for three different ventilation configurations and discuss the findings in the context of airborne transmission of COVID-19.

The paper is organized as follows: first, we introduce the underlying system of equations, the IB method, the concept of mask functions and sources/sinks as well as the definition of hard walls in this context. Next, we present 2 canonical reference cases for validating the code. Then, the performance of the newly developed DNSLABIB code is demonstrated in 1) the ventilation-induced emptying of a room of stale air, utilizing various ventilation setups, and 2) the emission of exhaled aerosols from respiratory activities such as speaking. Finally, we reiterate the main results and insights obtained in these simulations in light of the airborne infection risk.

\section{Theory and methods}
\subsection{Theoretical and numerical framework}
In the present work the focus is on low-speed, isothermal gas flows which can be modeled using the incompressible Navier-Stokes equations. Additionally, we study the transport of a passive scalar field representing the indoor CO$_2$ concentration. At the end of the paper, the airborne trajectories of a small number of Lagrangian particles is also studied assuming one-way coupling. The pressure-velocity coupling is based on the projection method~\cite{canuto2007spectral}, where the time integration is carried out using a 4th order explicit Runge-Kutta scheme. Additionally, the spatial derivatives in the momentum equation are discretized using 2nd order central differences in the skew-symmetric, energy conservative form (see e.g.~\cite{vuorinen2014implementation}). In the projection step, the pressure equation is solved in the Fourier space using the highly efficient fast Fourier transform (\texttt{fft}) method in MATLAB. The \texttt{fft} method is considered to be a key enabler for large scale CFD simulation in Matlab although it restricts the simulations to fully Cartesian, equispaced grids. The projection step is executed only once per time step in order to speed up the code. Based on our experience, this approximation has a negligible influence on the actual numerical solution. 

The governing equations for the fluid read
\begin{align}
    & \nabla \cdot \mathbf{u} = 0 \label{eq:NS1}, \\
    & \frac{\partial \mathbf{u}}{\partial t} + \nabla \cdot \left(\mathbf{u} \mathbf{u} \right) =
    -\nabla p + \nu \nabla^2 \mathbf{u} + \mathbf{S}(x,y,z) \cdot \frac{\mathbf{u}_{set}-\mathbf{u}}{\tau_f} + \mathbf{b} \label{eq:NS2} \\
    & \frac{\partial c}{\partial t} + \nabla \cdot \left(\mathbf{u} c \right) = \alpha_c \nabla \cdot \left[ \beta(x,y,z) \nabla c \right] + \mathbf{S}(x,y,z) \cdot \frac{c_{set} - c}{\tau_f} \label{eq:pstrans}\\
    & c_{|\beta = 0} = 0, \ \mathbf{u}_{|\beta = 0} = 0, \frac{\partial c}{\partial n}_{|\beta=0}=0  \label{eq:BCs},
\end{align}
where $\mathbf{u}$ is the velocity, $p$ is the pressure divided by the fluid density $\rho$, $\nu$ is the kinematic viscosity, $\mathbf{b}$ is a body force, $c$ is the passive concentration field and $\alpha_c$ is the diffusivity of the concentration. The binary mask function $\beta$ is used to mark the fluid phase ($\beta=1$) and the solid phase ($\beta=0$) respectively. On the no-slip walls, velocities are simply multiplied by $\beta$.

In Eqs.~\eqref{eq:NS1} and~\eqref{eq:NS2}, two additional terms appear: $\mathbf{S}(x,y,z) \left({\mathbf{u}_{set}-\mathbf{u}}\right)/{\tau_f}$, and $\mathbf{b}$. The latter term is a simple body force which is needed for pressure driven flows. The former is a forcing term adjusting the velocity to a target value at the flow inlets such as windows and ventilation ducts {\it{etc}}. This approach is needed since the present solver is periodic in contrast to non-periodic cases where the Dirichlet/Neumann boundary conditions for $\mathbf{u}$ and $p$ can be provided as usual. In brief, the term $\mathbf{S}(x,y,z) \left({\mathbf{u}_{set}-\mathbf{u}}\right)/{\tau_f}$ is formulated in order to establish the desired velocity $\mathbf{u}_{set}$ within the relaxation time scale $\tau_f$. By setting $S(x,y,z) = 1$ at the specific location, where the velocity must reach the value $\mathbf{u}_{set}$ and $S(x,y,z) = 0$ otherwise, the coordinate dependent mask function $\mathbf{S}(x,y,z)$ geometrically confines the momentum source to the targeted region of the geometry. A respective source term is also utilized in the scalar equation in order to investigate mixing.

\begin{figure}[t]
\centering
 \includegraphics[width=0.98\textwidth]{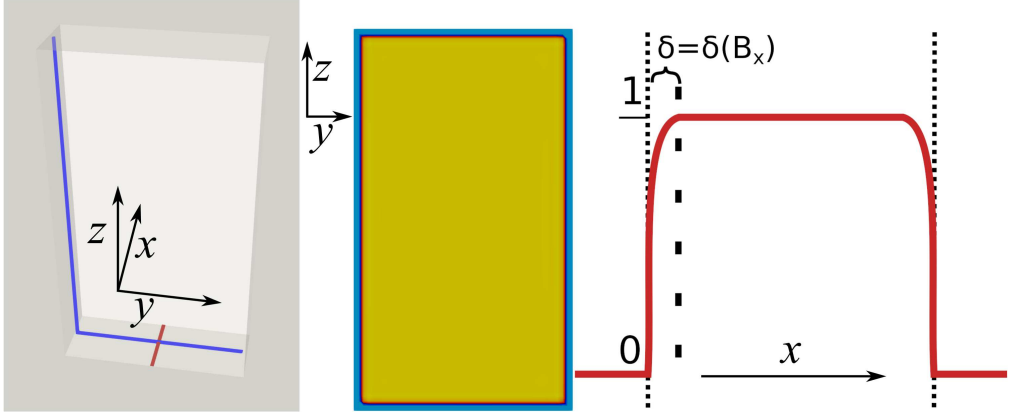}
 \caption{In the demonstration above, a window is located at a wall (left) and air is entering through the window. A plane, spanned by the blue lines, displays the inflow velocity field, constrained in place with the mask $S(x,y,z)$ (middle). Furthermore, the mask along the red line in the left-hand-side figure is plotted on the right.}
 \label{fig:fig1_1}
\end{figure}

A general mask function based on the hyperbolic tangent function is illustrated in Fig.~\ref{fig:fig1_1} and reads
\begin{align}
    S(x,y,x) = &\frac{1}{2} \left(1-\tanh(B_x \left[\frac{|x-x_c|}{W_x/2} - \frac{W_x/2}{|x-x_c|} \right]) \right) \cdot \nonumber\\
    &\frac{1}{2} \left(1-\tanh(B_y \left[\frac{|y-y_c|}{W_y/2} - \frac{W_y/2}{|y-y_c|} \right]) \right) \cdot \nonumber\\
    &\frac{1}{2} \left(1-\tanh(B_z \left[\frac{|z-z_c|}{W_z/2} - \frac{W_z/2}{|z-z_c|} \right]) \right) \label{eq:tanh} ,
\end{align}
where $x_c$, $y_c$ and $z_c$ are the volumetric center coordinates of the source region, while $W_x$, $W_y$ and $W_z$ are the window/inlet/ventilation duct dimensions of this region. This function obtains values in the range [0,1] and $B_{x,y,z}$ defines the smoothness of the transition between the two endpoints.
In the example of Fig.~\ref{fig:fig1_1}, a window is located at a wall (left) and an inflow of air with a constant velocity is forced with the mask function. The inflow velocity on a plane spanned by the blue lines smoothly decreases to zero at the windows edges (middle). The behavior of the mask function along the red line shows this continuous transition in more detail along the $x$-axis (right). Here, the width of the transition region, denoted by $\delta$, is highly dependent on the $B_x$ parameter in Eq.~\eqref{eq:tanh}.

\begin{figure}[h]
\centering
 \includegraphics[width=0.98\textwidth]{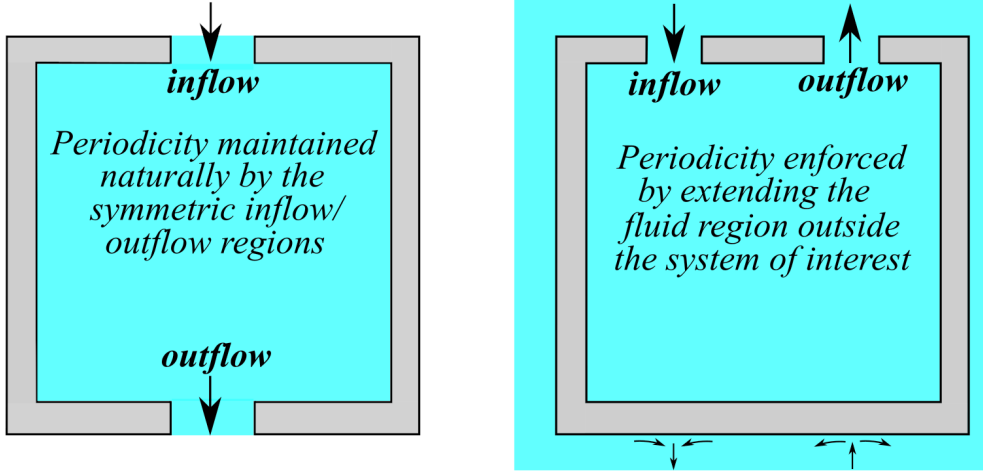}
 \caption{The spectral approach employed in solving the pressure equation in this work necessitates maintaining periodicity in the solution. This may be achieved naturally by a suitable inflow/outflow setup (left) or extending the fluid region beyond the system of interest (right).}
 \label{fig:fig1_2}
\end{figure}
Furthermore, the FFT approach in the pressure equation entails a periodic solution, which requires special consideration when creating inflow/outflow boundary conditions. In general, these can be imposed as follows in our code. Inflow through an opening (e.g. window or ceiling vent) requires either 1) modeling another window on the opposite side of the room from which the flow can exit (Fig.~\ref{fig:fig1_2} left), or 2) extending the fluid region around the room so that airflow can enter and leave the room to conserve mass (Fig.~\ref{fig:fig1_2} right). While option A is used in the cross-draught cases, option B is used in the ceiling ventilation case. We note that option B requires more computational resources since an extra flow passage needs to be modeled around the room.

Additionally, to model the subgrid scale effects and stabilize the flow, we utilize explicit filtering of the velocity and scalar fields at the end of each timestep using a 6th order filter. The filter is defined as follows 
\begin{align}
    \overline{\phi} = \phi + \Sigma_i \gamma_{x_i} \frac{\partial^6 \phi}{\partial x_i^6},
\end{align}
where the filter coefficient is chosen so that the Nyqvist frequency is zeroed in the Fourier space i.e. $\gamma_{x_i} = \frac{\Delta x_i^6}{\pi^6}$. The filter resembles a standard hyperviscosity term but avoids the cross-derivatives~\cite{johansen2005dust}.

\subsection{Solution of the pressure equation in Fourier space}
The pressure equation requires particular attention near the walls. In the conventional projection method, one obtains a $\mathbf{u}^*$ from the Navier-Stokes equation without a pressure gradient and then corrects $\mathbf{u}^*$ with the pressure gradient which is solved from the Poisson equation
\begin{align}
& \nabla \cdot \beta(x,y,z) \nabla p = \nabla \cdot \mathbf{u}^* \label{eq:Pressure} \\
    & \mathbf{u} = \mathbf{u}^* - \nabla p 
\end{align}

\noindent In the expression above, the mask function $\beta$ is simply used to implement the Neumann boundary condition directly to the Laplacian operator $\nabla \cdot \beta(x,y,z)\nabla p$. The Fourier transform of this equation can not be directly determined. However, by adding and subtracting 1 from $\beta$, one has $\nabla \cdot \beta(x,y,z)\nabla p= \nabla \cdot (1+\beta(x,y,z)-1)\nabla p$. Finally, the pressure equation can be recast into the following equation where the left hand side operator now has a well-defined Fourier transform 
\begin{align}
& \Delta p^{k+1} = \nabla \cdot \left( 1-\beta(x,y,z) \right) \nabla p^k + \nabla \cdot \mathbf{u}^* \label{eq:Pressure}.
\end{align}
\noindent We iterate equation~\eqref{eq:Pressure} $n$ times evaluating the right hand side of the equation by central differences from the previous available value of pressure ($p_k$) and the non-solenoidal velocity $\mathbf{u}^*$ acquired from the momentum equation. Generally, the equation converges very quickly and a hard-coded value $n=4$ is used here. The velocity field is then corrected as $\mathbf{u} = \mathbf{u}^* - \nabla p$, utilizing the converged solution for the pressure to yield the solenoidal field satisfying Eq.~\eqref{eq:NS1}. 

\subsection{Lagrangian particles}
Proceeding to Lagrangian particles, the equations of motion (EoM) read
\begin{align}
 & \frac{d \mathbf{u}_p}{dt} = \frac{C_d}{\tau_p}\left(\mathbf{u}_f - \mathbf{u}_p \right) + \mathbf{g}, \\
    & \tau_p = \frac{\rho_p d^2}{18 \rho_f \nu}, \\
& Re_p = \frac{|\mathbf{u}_f-\mathbf{u}_p|d}{\nu}, \\
    & C_d = 1 + \frac{1}{6}Re_p^{2/3} ,
\end{align}
where $Re_p$ is the particle Reynolds number, $\tau_p$ is the particle settling time scale, describing the delay of the particle in adjusting to altered flow conditions, and $C_d$ refers to the drag coefficient of a particle. The subscripts $p$ and $f$ indicate particle and fluid properties, respectively, $g$ is the gravitational force, $d$ is the particle radius and $\rho$ is the (bulk) density. These equations  describe the trajectory of a particle which experiences the effect of gravity as well as the drag imposed by the surrounding fluid. Since the particle relaxation time scale $\tau_p$ is generally small for particles of interest here (micrometer scale), the particle equations of motion are solved using the implicit Euler method, enabling larger time steps. Two key quantities from the EoM above are the particle terminal velocity $v^{*}=g \tau_p$ as well as the particle sedimentation time from height $h$ expressed as $\tau_s = h/v^{*}$. For $h=1.5$ m and $d$=5/10/30 $\mu$m solid particles $\tau_s$ is approximately 36/9/1 min respectively if the ambient airflow is non-existent. In practice this simple analysis (see e.g. \cite{vuorinen2020modelling}) indicates that particles with sizes up to 100 $\mu$m can traverse significant distances and thus they pose a risk of also being inhaled. We further address this aspect at the end of this paper.    

\subsection{Overview of the DNSLABIB code}
\begin{figure}[t]
\centering
 \includegraphics[width=0.68\textwidth]{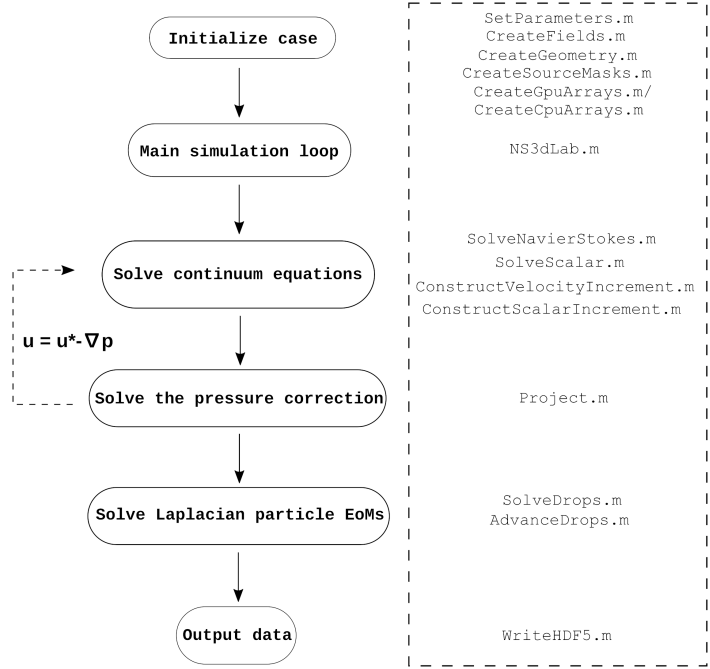}
 \caption{A flow chart illustrating the code execution pipeline.}
 \label{fig:fig1}
\end{figure}

As stated earlier, we implement a numerical code to solve the governing equations~\eqref{eq:NS2},~\eqref{eq:pstrans} and~\eqref{eq:Pressure} using the MATLAB language. Accordingly, no code compilation nor external dependencies are required and the supported platforms currently include Windows, Linux and macOS. Our open-source code is freely available and the structure of the program is illustrated in Fig.~\ref{fig:fig1}. The user initializes a case and controls the subsequent simulation principally via the \texttt{"SetParameters.m"}, \texttt{"CreateFields.m"}, \texttt{"CreateGeometry.m"}, \texttt{"CreateSourceMasks.m"} and \texttt{"InitializeDrops.m"} scripts. In \texttt{"SetParameters.m"}, the user provides the necessary information regarding the simulation geometry, fluid properties and data outputting. Importantly, the user also specifies the maximum Courant number, as the implementation utilizes dynamic time stepping (in \texttt{"AdjustDt.m"}). In \texttt{"CreateGeometry.m"}, the user specifies the obstacles in the flow domain by forming the $\beta (x,y,z)$ field with primitive shapes. Then, in \texttt{"CreateSourceMasks.m"}, the user specifies the location of the sources/sinks. The user also decides on whether the simulation is run on CPUs or GPUs, leading to calls to either \texttt{"CreateCpuArrays.m"} or \texttt{"CreateGpuArrays.m"}. Finally, the number, size and location of Lagrangian particles is assigned in \texttt{"InitializeDrops.m"}, after which the simulation may be launched by running \texttt{"NS3dLab.m"}. This main simulation loop calls various other functions designed to solve the fluid and particle equations which we introduced above. \texttt{"SolveNavierStokes.m"} (\texttt{"SolveScalar.m"}) implements the explicit RK4 time integration, updating the convective and diffusive terms in the Navier-Stokes (passive scalar transport) equations via calls to \texttt{"ContructVelocityIncrement.m"} (\texttt{"ContructScalarIncrement.m"}). Finally, the projection step is performed in \texttt{"Project.m"}. In an analogous manner, \texttt{"SolveDrops.m"} advances the particle trajectories and velocities in time by constructing the terms in the equations of motion in \texttt{"AdvanceDrops.m"}.
Based on the settings provided by the user in \texttt{"SetParameters.m"} the main loop issues calls intermittently to \texttt{"writeHDF5.m"} to output simulation data.
Since for large systems containing tens or hundreds of millions of cells and a vast number of time steps, the fluid data may routinely reach extreme sizes, and therefore, special consideration must be given to the data format. Currently, the code can be set to output the fluid data in HDF5 compliant format along with the associated XMDF metadata file or in the MATLAB native format (.mat). Additionally, the Lagrangian particle data is outputted in raw text data. In our experience, visualizing large fluid data sets in HDF5 format via external software, such as ParaView, yields superior read and rendering performance to many other options available. For standard users with more modest data outputting requirements, the flow quantities, such as the velocity field, pressure, the passive concentration (and their respective time averages) can also be forwarded to a \texttt{.mat} file for quick access and analysis.

\subsection{Advantages and limitations of DNSLABIB}
As noted, the main objective of the paper is to develop an efficient GPU-compatible code in MATLAB for performing scale-resolved 3D CFD simulations. Before proceeding to results, the expected advantages and limitations of DNSLABIB should be mentioned. The expected advantages of DNSLABIB include: 

\begin{itemize}
    \item Potentially high performance, specifically for large systems, due to the GPU implementation. The primary contributors to the observed performance include avoiding loops (vectorization) as well as the efficient {\texttt{gpuArray}} structure and {\texttt{fft}} function in MATLAB  
    \item Avoiding the use of a supercomputer. 
    \item Ease of use in configuring and executing a case for simple geometries with solid objects.
    \item Scale-resolved simulations present the opportunity to capture physical flow features such as shear-driven turbulent flow, mixing, and flow recirculation zones.   
\end{itemize}
\noindent In contrast, the main limitations of the present DNSLABIB implementation include: 
\begin{itemize}
    \item Local mesh refinement is presently not possible due to the incorporation of \texttt{fft} for the pressure equation. The mesh resolution needs to be uniform. 
    \item Since mesh refinement is not possible, only vents and windows with a relatively large diameter can be resolved (e.g. over 20 computational cells per diameter) at the moment.
    \item Wall models are currently not implemented in the code and one needs to rely on constant grid resolution even at the walls.
    \item The obstacles are presented as simple blocks.  
    \item Thermal sources/sinks are not presently accounted for although they are known to be of high importance in indoor airflow configurations.
\end{itemize}

In the present paper, the main focus is in understanding the GPU compatibility of the present immersed boundary approach in MATLAB. Here, wall functions are not in the focus but, instead, we aim at resolving the shear-driven flows as well as possible on uniform grids. Commonly, the absence of wall models is considered detrimental to the solution accuracy in wall-turbulence driven, high Reynolds number flows, especially on coarse computational grids. Here, we carry out a sensitivity assessment on Reynolds number effects and discuss the cases $Re=5000$ (window airflow 0.1m/s) and $Re=50000$ (window airflow 1m/s) to better understand how $Re$ affects the ventilation rate when the airflow velocity changes. In the present cases, near-wall air velocities are rather low speed on the order of $\sim 0.01-0.1$ m/s so that the length scales of the viscous wall layer scales $y+<5-10$ are mostly reached. For $Re=50000$, 88\% of the computational cells have $y+<10$. For $Re=5000$ all near-wall cells have $y+<10$ while $\approx$98\% of the near-wall cells have $y+<5$.

\begin{comment} First, we note that indoor turbulence is commonly driven by free shear layers. Thus, near-wall turbulence production is not expected to dominate the airflow structure. Furthermore, typical indoor air velocities are very low speed in the order of $\sim 0.01-0.1$m/s so that the length scales of the viscous wall layer thickness (e.g. $y+ \sim 5-10$) may be commonly reached on a grid resolution of a few centimeters. Finally, in real indoor geometries the uncertainty produced by solid obstacles (e.g. furniture, items on the desk etc) is in the order of $\sim 0.01-1$m where a typical grid size in DNSLABIB is $\approx 0.02$ m herein. Such solid obstacles produce pressure loss thereby having a profound impact on the turbulent mixing as well. Considering such complex circumstances, formulating consistent wall functions is no. In the following, we discuss the performance of DNSLABIB in a number of airflow setups with solid obstacles. 
\end{comment}

\section{Results}
Next, the performance of DNSLABIB is explored by introducing several simulation cases of increasing complexity. First, the code is validated in two canonical reference cases: the laminar channel flow and vortex shedding due to a rectangular body mounted into the channel. Then, we further validate the code against a scale-resolved simulation of an indoor ventilation setup performed in OpenFOAM. A number of indoor ventilation setups are then examined to study the removal of stale air from the room and assess the infection risk in these configurations in the context of COVID-19.
\subsection{Validation and benchmarking}
\subsubsection{2D laminar flows}
\begin{figure}[t]
\centering
 \includegraphics[width=0.95\textwidth]{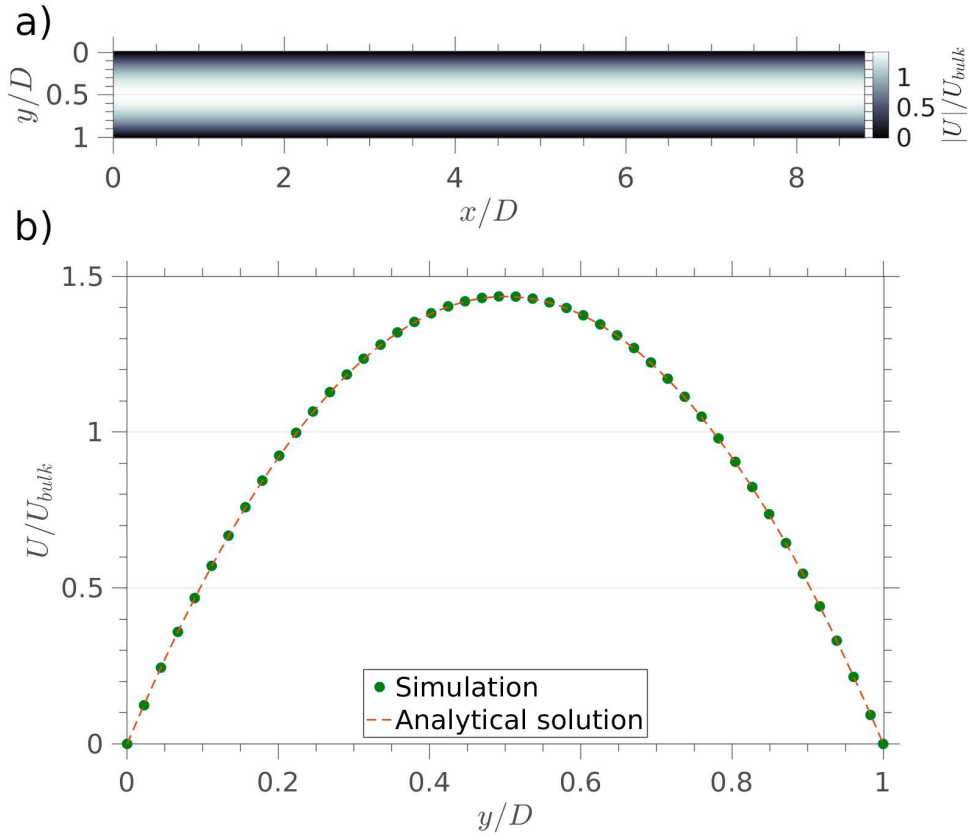}
 \caption{Velocity in the laminar channel flow (a) and matching the analytic, parabolic velocity profile (b).}
 \label{fig:fig7}
\end{figure}

\begin{table}[h]
    \centering
    \begin{tabular}{c|c|c} 
         Property & Channel & Channel \& Cubic obstacle  \\ \hline
         Channel width ($D$) [m]  & $2\pi$ $\approx$ 6.28 & $2\pi$ $\approx$ 6.28 \\
         Wall width [m]  & $0.05 \cdot D \approx 0.31$ & $0.05 \cdot D \approx 0.31$ \\
         $x$-dimension ($L_x$) [m] & $8.8 \cdot D$ $\approx$ 55.3 & $8.8 \cdot D$ $\approx$ 55.3  \\
         $y$-dimension ($L_y$) [m]  & $1.1 \cdot D$ $\approx$ 6.91 & $1.1 \cdot D$ $\approx$ 6.91 \\
         Obstacle edge length ($h$) [m] & -- & $D/6 \approx 1.05$ \\
         Grid & 1584 $\times$ 198 & 1584 $\times$ 198 / 3168 $\times$ 396 \\
         $Re$ & 500 & 84 (at obstacle) \\
    \end{tabular}
    \caption{The simulation parameters for the flow in a channel and vortex shedding cases.}
    \label{tab:tab1}
\end{table}
First, a laminar channel flow driven by a pressure gradient is examined to test the IB method. The simulation parameters for this case are detailed in Tab.~\ref{tab:tab1}. The flow is initialized by setting a body force (acceleration) in the $x$-direction $\mathbf{b} = {\tilde{\mathbf{b}}/\rho} = (b_x,0,0)$. In this simple channel flow, the Navier-Stokes equations imply an analytical velocity profile across the channel $U(y) = (-b_x / 2\nu) y^2 + (b_x D / 2\nu) y$ which translates to a bulk velocity of $U_{b}=\frac{1}{D}\int_0^{D} U(y) dy = \frac{b_x D^2}{12 \nu}$. Since the Reynolds number is $Re = U_{b} D / \nu = b_x D^3 / 12 \nu^2$, fixing the ratio $b_x / \nu$ fixes the Reynolds number, which is set to $Re = 500$ here. As noted in Fig.~\ref{fig:fig7} b), the present simple example indicates that the analytical parabolic velocity profile is recovered in the simulation with the relative error remaining below $10^{-5}$ .

\begin{figure}[h!]
\centering
 \includegraphics[width=0.95\textwidth]{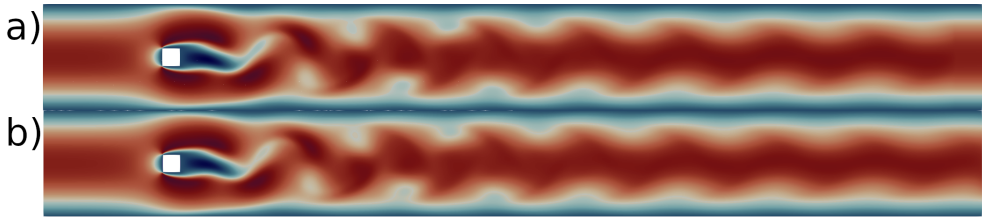}
 \caption{A cube in a channel flow. The $x$-component of velocity obtained in DNSLABIB (a) and OpenFOAM (b).}
 \label{fig:fig9_2}
\end{figure}
Next, as depicted in Fig.~\ref{fig:fig9_2} a), a cube is placed at the centre line of the channel. The purpose of the test case is to demonstrate the performance of DNSLABIB when the wake interacts dynamically with the channel walls, resulting in vortex shedding. The cube is located at $(1.2D, 0.5D)$ and the side length is $h=D/6$ while the rest of the geometry remains unaltered from the previous case discussed. Setting $\Delta x=\Delta y$ we explore the case utilizing two uniform mesh densities: $\Delta x = h/30$ and $h/60$. A parabolic velocity profile at the left-hand-side of the channel is forced, with no-slip conditions at the channel walls and a maximum value $U_{max}$ in the center of the inflow region ($y = D/2$). This translates to the obstacle Reynolds number of $Re_{ob} = U_{b} \cdot h / \nu = \frac{2}{3} U_{max} \cdot h / \nu \approx 84$ being in the unsteady vortex shedding regime. The subsequent Kármán vortex street is examined in more detail in Fig.~\ref{fig:fig9_2}, where the $x$-component of the velocity field is examined in steady-state in both DNSLABIB (a) and OpenFOAM (b). The reference OpenFOAM simulations are carried out with the standard incompressible PIMPLE-solver utilizing a second order accurate backward method in time, the linear (corrected) interpolation for the convection terms, and linear central differencing for the diffusion terms (see e.g. \cite{peltonen2019large} for a similar approach). In contrast to DNSLABIB, OpenFOAM boundary conditions are provided in a standard manner using Dirichlet/Neumann conditions. 

As expected, both approaches indicate periodic vortex shedding at a distinct oscillation frequency $f$. For the fine grid DNSLABIB simulations, the Strouhal number of the oscillation  $St=fh/U_{b}\approx 0.2094$ while the mean drag coefficient $C_D\approx 3.05$. The results are in excellent agreement with the respective values computed in OpenFOAM on the same grid resolution with $St\approx 0.2094$ and $C_D \approx 3.09$. 
\begin{table}[h]
    \centering
    \begin{tabular}{c|c|c|c} 
         Code & Cells per obstacle size & Strouhal number (St) & Drag coefficient ($C_D$)  \\ \hline
         DNSLABIB & 30 $\times$ 30 & 0.2097 & 3.12 \\
         DNSLABIB & 60 $\times$ 60 & 0.2094 & 3.05 \\
         OpenFOAM & 30 $\times$ 30 & 0.192 & 3.21 \\ 
         OpenFOAM & 60 $\times$ 60 & 0.2094 & 3.09
    \end{tabular}
    \caption{A cube in a channel flow: comparison of the Strouhal number and the drag coefficient obtained by DNSLABIB and OpenFOAM on two different grid resolutions.}
    \label{tab:tab2}
\end{table}

\subsubsection{3D turbulent flow indoors}
In the following, airflow is simulated in a more challenging validation case at $Re=5000$ involving a large indoor space (8 m x 8 m x 3 m along $x$-, $y$- and $z$-axis, respectively) with open windows. 
The space is divided into two larger rooms and a corridor (3 rooms in total) which are connected by doorways. A cross-draught is then generated as air enters and exits through the windows, ventilating the room. In addition to the $Re=5000$ with window airflow peak velocity $U_{in}=0.1$m/s, we also investigate a higher inflow velocity of $U_{in}$=1m/s corresponding to $Re=50000$. The main validation is based on the $Re=5000$ case which is also rather challenging in terms of fluid dynamics while the $Re=50000$ case is investigated to better understand the $Re$ sensitivity of DNSLABIB and to show that the code stays numerically stable at extreme Reynolds numbers as well. Both of these cases are turbulent and hence non-trivial.   

\begin{figure}[t]
\centering
 \includegraphics[width=0.94\textwidth]{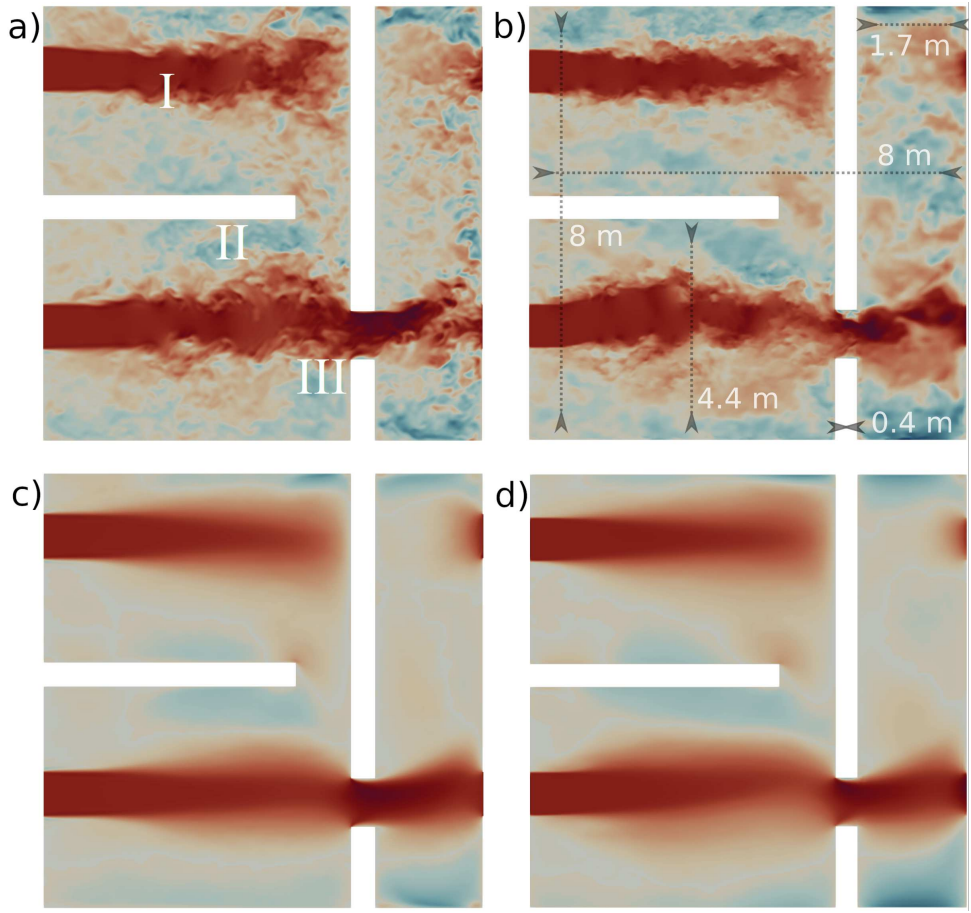}
 \caption{Instantaneous $x$-component of velocity at the midcut plane using DNSLABIB (a) and OpenFOAM (b). The time-averaged velocity fields obtained in DNSLABIB and OpenFOAM are visualized in (c) and (d) respectively. \uppercase\expandafter{\romannumeral 1\relax} Kelvin-Helmholtz instability, ~\uppercase\expandafter{\romannumeral 2\relax} recirculation zones and ~\uppercase\expandafter{\romannumeral 3\relax} flow acceleration.}
 \label{fig:fig4_23}
\end{figure}

Fig.~\ref{fig:fig4_23} illustrates the setup in more detail for $Re=50000$. The $Re=5000$ case would remain qualitatively very similar with 10$\times$ slower timescales due to the 10-fold lower velocity scales. The midcut plane displaying the cross-section of the room is portrayed here at $z=1.5$ m. The $x$-component of the instantaneous velocity field is presented comparing DNSLABIB (a) and OpenFOAM (b) at the midcut. The panels (c) and (d) present the temporally averaged $x$-component of velocity for DNSLABIB and OpenFOAM respectively. Here, red color implies positive value for the velocity component, while blue suggests negative values. First, the mean velocity data in c) and d) indicate a good agreement. The key physical phenomena of the turbulent flows are marked  in a) as follows. ~\uppercase\expandafter{\romannumeral 1\relax}: The shear layer generated turbulence exhibits a Kelvin-Helmholtz instability a few window widths from the window (see also Fig.~\ref{fig:fig4_31}). \uppercase\expandafter{\romannumeral 2\relax}: Negative velocities and formation of recirculation zones are observed next to the walls aligned with the flow direction. \uppercase\expandafter{\romannumeral 3\relax}: The flow accelerates at the more narrow doorways. These physical phenomena are well known and  expected and it is therefore crucial that they are correctly captured by DNSLABIB and in qualitative agreement with the OpenFOAM code.  

\begin{figure}[t]
\centering
 \includegraphics[width=0.98\textwidth]{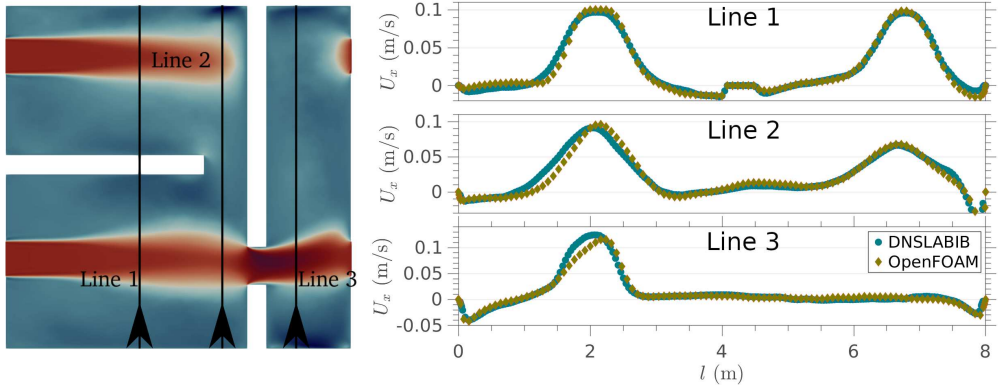}
 \caption{The time averaged $x$-component of velocity obtained in DNSLABIB and OpenFOAM along three sampling lines (b) for $Re=5000$ (low ventilation rate). The sampling lines are displayed in (a) at the midcut of the simulation geometry and the arrows indicate the plotting direction (from 0 to 8m). The agreement appears very good.}
 \label{fig:fig4_221}
\end{figure}

Fig.~\ref{fig:fig4_221} a) details the location of three sampling lines on the midcut plane, along which the x-velocity component is interpolated and compared between the two computational tools in b) for the $Re=5000$ case.  The comparison indicates good agreement between DNSLABIB and the OpenFOAM simulations. 
Notably, the comparison in the present flow setup is computationally rather demanding because the flow is highly transitional and the window width is large compared to the room dimensions so that the walls are relatively close to the shear layers. Considering these aspects, the present results for DNSLABIB at $Re=5000$ can be considered to be in very good agreement with OpenFOAM. For $Re=50000$, the agreement is still  satisfactory. We remind that for the lower Reynolds number all near-wall cells have $y+<10$ while 98\% of them are below $y+<5$. For the higher Reynolds number approximately 88\% of the cells have $y+<10$. Hence, the present results provide numerical evidence that, in particular for the $Re=5000$ case, the mean velocity gradients are well resolved all the way to the walls.  

\subsection{DNSLABIB performance on a GPU}
\begin{figure}[h!]
\centering
 \includegraphics[width=0.88\textwidth]{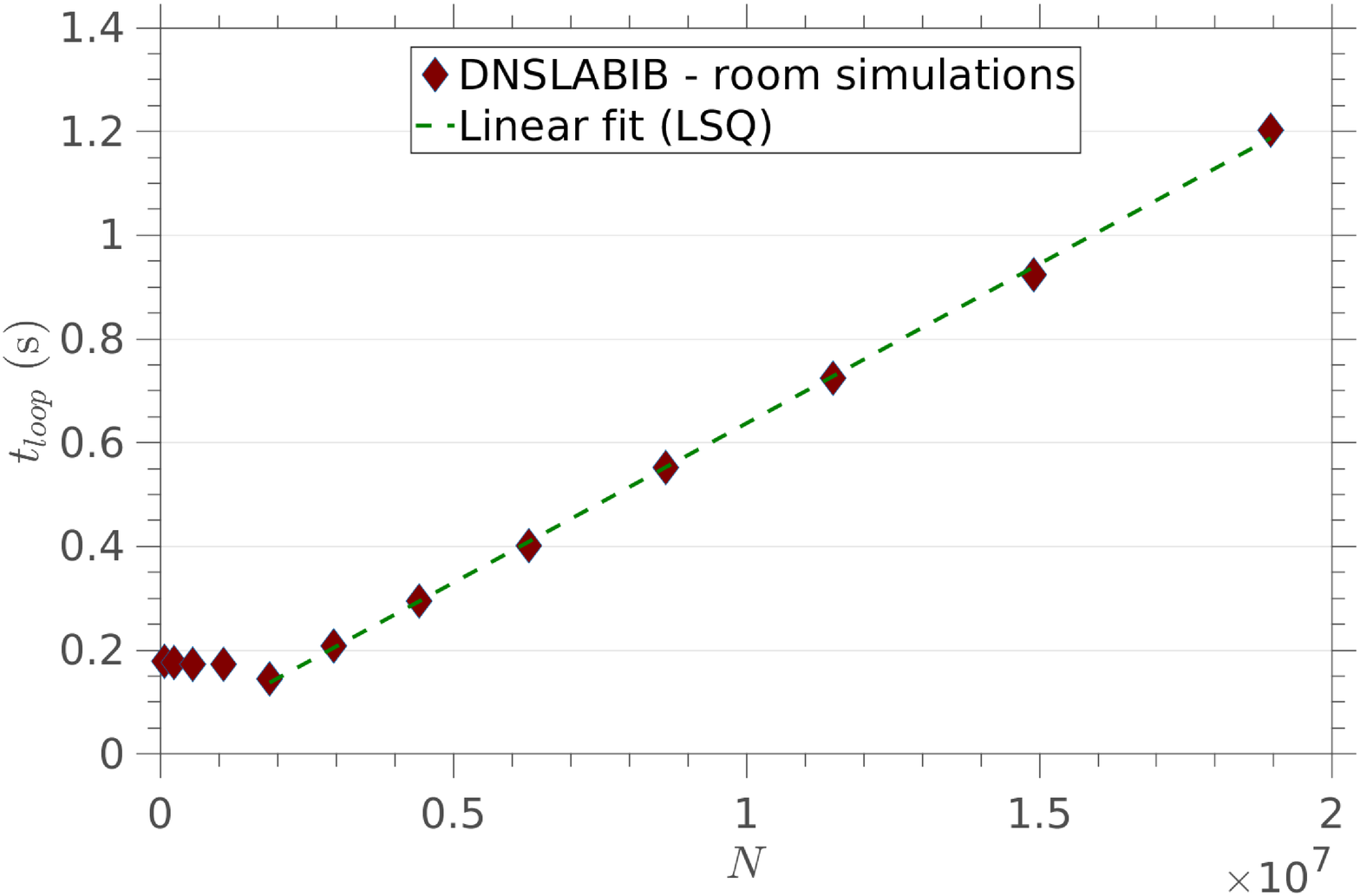}
 \caption{The computational time required as a function of the system size plotted for the 3RW case at $Re=50000$ (high ventilation rate). The relationship seems linear, which is ideal, indicating there is no extraneous communication overhead during the simulation. The device memory (NVIDIA Tesla V100) inhibits further benchmarking beyond $N = 2 \cdot 10^7$.}
 \label{fig:fig4}
\end{figure}
Next, we discuss the simulation time for the three-room configuration emphasizing the performance of DNSLABIB in comparison to OpenFOAM. The number of total computational cells for the DNSLAB case is approximately $33.3 \cdot 10^6$ while the OpenFOAM case contains $\sim 16.2 \cdot 10^6$ cells. The DNSLABIB run for a reference room simulation takes approximately 48 hours on a single NVIDIA Tesla A100 GPU. In contrast, the parallel OpenFOAM simulation, executed on 1040 CPU cores (Intel Xeon Gold 6230), consumes approximately 155 hours of computational time for the respective simulated physical time. Not only is the DNSLABIB run by a factor of 3x faster than the OpenFOAM run with almost double the computational cell count, DNSLABIB can run on a more lightweight platform containing a single GPU, avoiding using a supercomputer. 

A systematic DNSLABIB scaling test for the reference room case is shown in  Fig.~\ref{fig:fig4} detailing the computational time as a function of the mesh size as the mesh is refined. The computational time is defined as the execution time of a single (NVIDIA Tesla V-100) GPU to reach a simulation time of 120 seconds (starting from $t_0 = 0$ s). The relationship is linear which is an ideal result indicating no superfluous overhead is generated in simulations involving a larger number of computational cells. This result is reasonable, since the simulation data is completely contained within the VRAM of the GPU in our benchmark cases and thereby any degree of extraneous communication overhead should be avoided.

\subsection{Three indoor airflow configurations}
\subsubsection{Room setups and overview of the airflow}
\begin{figure}[t]
\centering
 \includegraphics[width=0.96\textwidth]{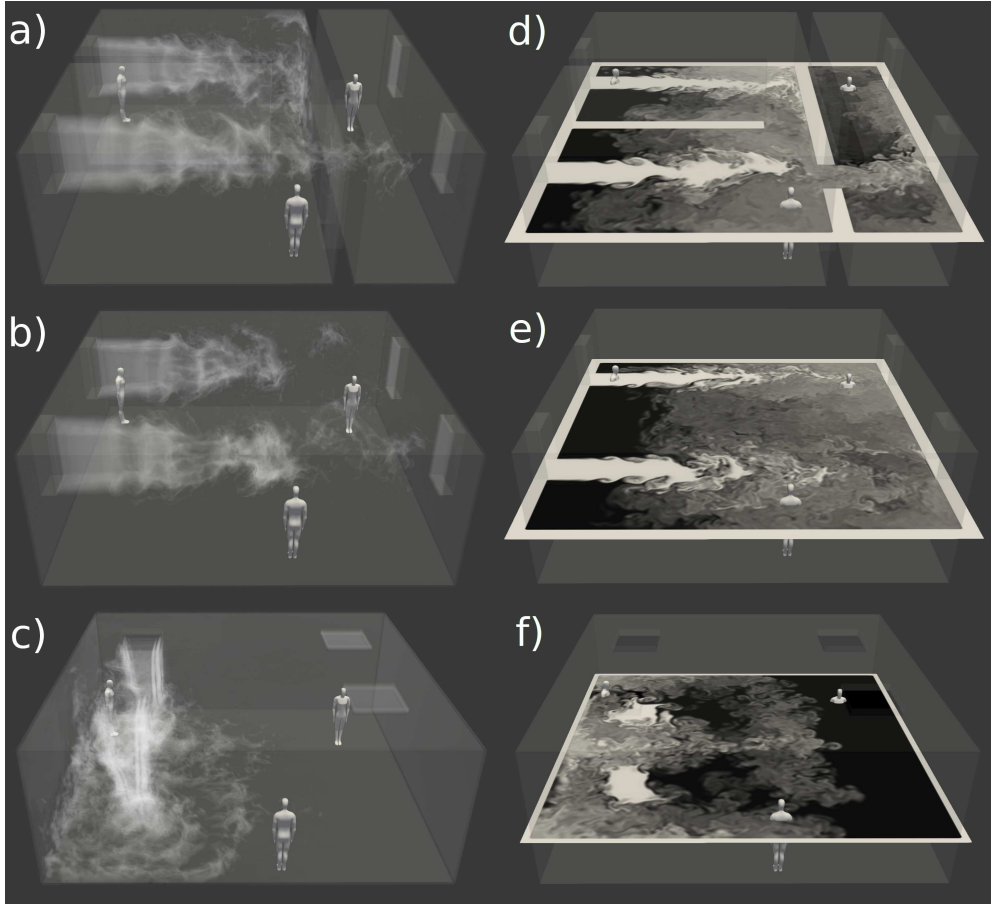}
 \caption{Three ventilation setups with an identical airflow rate. a) Cross-draught via windows and the space is divided to 3 smaller rooms (3RW). b) Cross-draught via windows without room-dividers (1RW). c) Vents located at the ceiling without room-dividers (1RV). d)-e) display slices of the instantaneous CO$_2$ fields for the respective cases.}
 \label{fig:fig3}
\end{figure}

 In addition to the previously validated case, Fig.~\ref{fig:fig3} displays two additional ventilation setups with the white cloud portraying the entering fresh airflow which is assumed to be isothermal. Here, the studied rooms are empty even though simple block shaped furniture can be easily included in DNSLABIB.  
Fig.~\ref{fig:fig3} a) corresponds to the original 3-room case while b) shows a case where the room dividers have been removed and the resulting space becomes one single open room. In panel c), the room dividers remain removed but the ventilation airflow is now generated by vents located at the ceiling as the windows are closed. In reality, the incoming airflow is commonly directed by grids to enhance mixing, break strong air currents and thereby enhance the indoor airflow comfort as well. Here, however, the inflow is simply modeled as plain air jets. The parameters applied in these cases are detailed in Tab.~\ref{tab:tab3} and in the following discussion, the cases are referred to as 3RW (3 rooms with windows), 1RW (1 room with windows) and 1RV (1 room with vents), respectively. For consistency, we have matched the inflow rate of air in these cases.

\begin{table}[t]
    \centering
    \begin{tabular}{c|c|c|c} 
         Property & 1 room (windows) & 3 rooms (windows) & 1 room (vents)  \\ \hline
         Case abbreviation & 1RW & 3RW & 1RV \\ 
         $x$-dimension ($L_x$) [m] & 8 & 8 & 8  \\
         $y$-dimension ($L_y$) [m]  & 8 & 8 & 8 \\
         $z$-dimension ($L_z$) [m]  & 3 & 3 & 3 \\
         Grid (fine) & 430 $\times$ 430 $\times$ 180 & 430 $\times$ 430 $\times$ 180 & 450 $\times$ 450 $\times$ 200 \\
         Kin. viscosity [m/s$^2$]  & $1.6 \cdot 10^{-5}$ (air) & $1.6 \cdot 10^{-5}$ (air) & $1.6 \cdot 10^{-5}$ (air) \\
         $U_{in}$/$U_{out}$ [m/s]  & 1.0/0.1 & 1.0/0.1 & 1.0/0.1 \\
         $A_{wind}$ [m$^2$]  & 1.2 & 1.2 & -- \\
         $A_{vent}$ [m$^2$]  & -- & -- & 1.2 \\
         Window count & 4 & 4 & -- \\
         Vent count & -- & -- & 4 \\
         $Re$ & 50000 / 5000 & 50000 / 5000 & 50000 / 5000 \\
    \end{tabular}
    \caption{The simulation parameters for the three room configurations for the two Reynolds numbers. }
    \label{tab:tab3}
\end{table}
Panels d)-e) in Fig.~\ref{fig:fig3} also display the instantaneous spatial patterns in the CO$_2$ concentration field for the three setups at the midcut plane at $t=20$ s for 3RW (d), 1RW (e) and 1RV (f) cases respectively ($Re=50000$). Here, the black (white) color designates areas of stale (fresh) air. For instance, in e) and f), numerous pockets of stale air appear which remain stagnant and poorly ventilated. Some of these pockets are expected and intuitive (room corners) while some are less intuitive such as the boundary of the wall separating the two windows in e). The formation of stagnation zones is expected based on known fluid dynamics and flow recirculation near the room corners.  Furthermore, turbulent airflow affects the mixing of the fresh and stale air. We note that qualitatively very similar results are observed for the cases with $Re=5000$ but the flow simply evolves 10 times slower due to the lower window airflow velocity (not shown herein for brevity). From the viewpoint of infection risk, strong variance in the CO$_2$ content of a room also indicates a potential spatial variance in the infection risk of an airborne disease. While indoor CO$_2$ measurements have recently been employed as a proxy to monitoring the virus concentration during the COVID-19 pandemic~\cite{villanueva2021assessment,kitamura2021co2,poza2021indoor,chen2021recommendations}, one could argue that the information provided by CO$_2$ sensors can yield misleading output if such local variances are not quantified while designing the measurement setup. The observations highlight the importance of accounting for the uncertainty resulting from the geometric features of indoor spaces. Gaining a complete three dimensional description of the airflow characteristics is typically only accessible via scale-resolved CFD simulations. In practice, personal CO$_2$ meters offer real time air quality monitoring at the location of any individual. 

\subsubsection{Ventilation characteristics}
A commonly used engineering metric for the ventilation rate is the air changes per hour ([ACH]=1/h) defined as
\begin{equation}
    \mathrm{ACH} = \frac{\dot{V}}{V} ,
\end{equation}
\noindent where $V$ ($[V]$=m$^3$) is the room volume and $\dot{V}$ is the volumetric airflow into the room ($[\dot{V}]$=m$^3$/h). In practice, ACH depends on the room airflow details, heat sources and geometrical features. ACH value can be measured by CO$_2$ measurements~\cite{schibuola2021high}. Here, we estimate ACH as follows ~\cite{bartzanas2007analysis}. A room is initially saturated with a relatively high CO$_2$ content stale air (here: 1000 ppm). Then, fresh outdoor air with a low CO$_2$ concentration (here: 400 ppm) is released into the room via windows or ventilation ducts. The mean CO$_2$ concentration as a function of time can then be monitored to yield the ACH value. 

\begin{figure}[t]
\centering
 \includegraphics[width=0.94\textwidth]{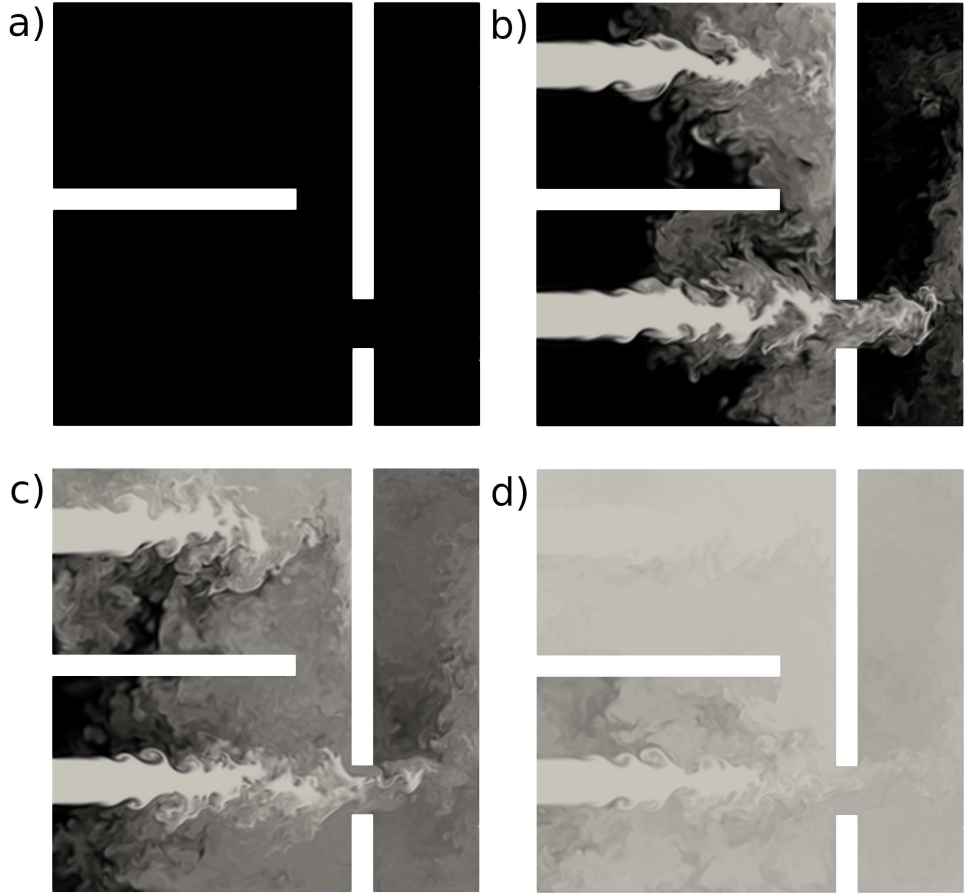}
 \caption{The substitution of stale air i.e. 1000 ppm CO$_2$ concentration (black) with fresh air i.e. 400 ppm CO$_2$ concentration (white) in the 3RW case using DNSLABIB at $U_{in}=1m/s$ ($Re=50000$, high ventilation rate). Midcut planes of CO$_2$ concentration at a) $t=0$ s, b) $t=20$ s, c) $t=60$ s and d) $t=180$ s.}
 \label{fig:fig4_31}
\end{figure}

Fig.~\ref{fig:fig4_31} demonstrates the CO$_2$ concentration in the 3RW simulation ($Re=50000$), where the stale air is gradually displaced by clean air over time. The window jets are noted to either impinge on the opposite wall in the back room or alternatively exit almost directly through the doorway to the corridor. Regions of lower ventilation performance can emerge near wall corners within re-circulation areas (e.g. the region in close proximity of the window at lower left corner of the frame). For the 1RW case, the absence of the additional walls imply that the window jets exit through the opposite windows with relatively less mixing of stale and clean air. 

\begin{figure}[h!]
\centering
 \includegraphics[width=0.98\textwidth]{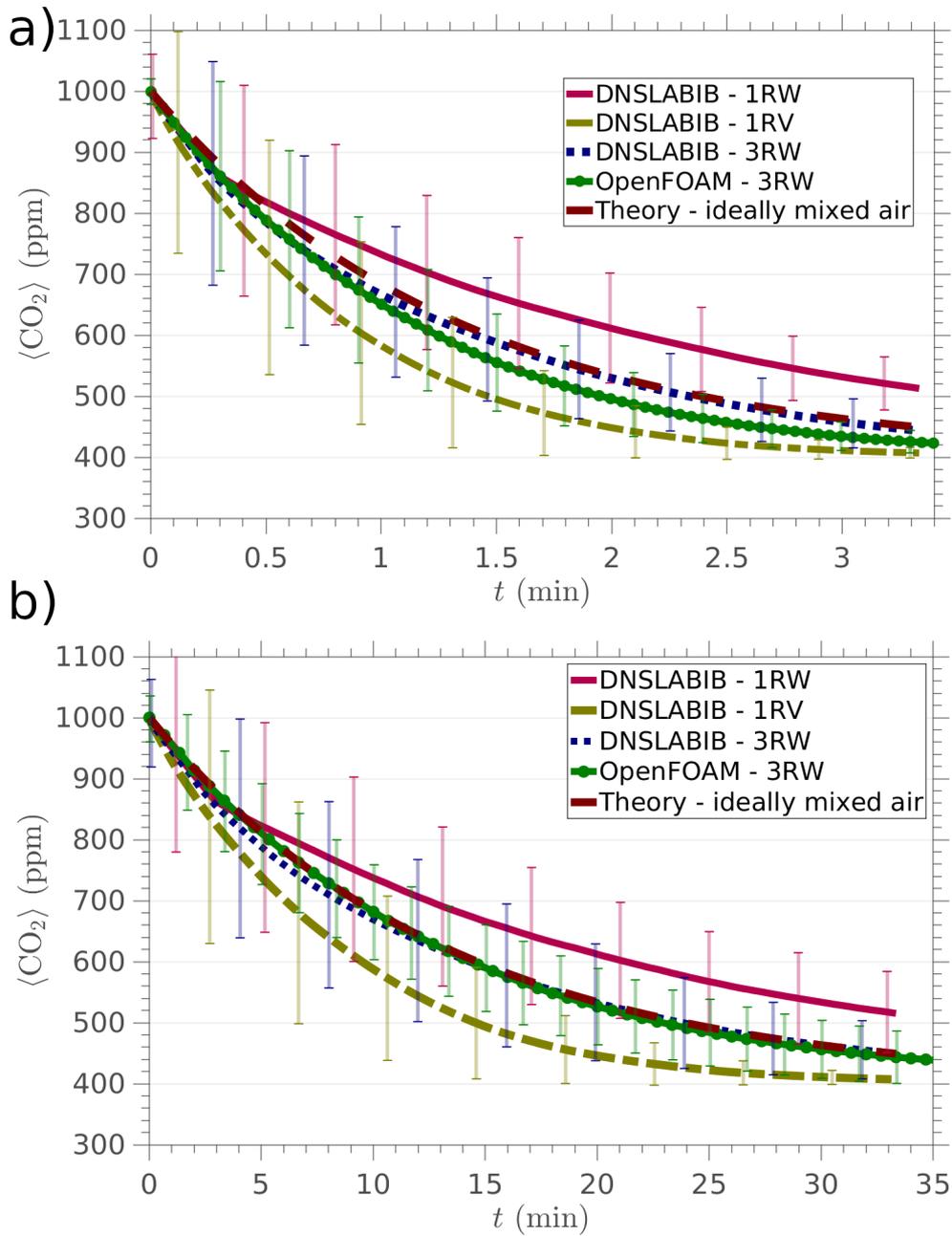}
 \caption{The time-evolution of the mean carbon dioxide content in each of the ventilated room scenarios presented earlier for the high ventilation rate, $Re=50000$ (a) and the low ventilation rate, $Re=5000$ (b). The concentration in the room, which is initially saturated with stale air, decreases in an almost exponential manner. Clear deviations from the theoretically derived behavior is observed.}
 \label{fig:fig4_3}
\end{figure}

The mean CO$_2$ concentration can be monitored to yield the actual ACH value of each ventilation setup which differs from the theoretical ACH value. In Fig.~\ref{fig:fig4_3}, the mean $\mathrm{CO}_2$ content as a function of simulation time $t$ is plotted for the studied cases. Panel a) corresponds to simulations at $Re=50000$ and b) to $Re=5000$. The standard deviation is plotted as well with the error bars to indicate the uncertainty of the $\mathrm{CO}_2$ distributions for each case. Simultaneously, the analytical expression $400 \cdot \exp(-\mathrm{ACH} \cdot t) + 600$ based on the ventilation theory of perfectly mixed air is displayed. The theoretical ACH coefficient is determined as the ratio of the flow mass fluxes involved and the room volume for both $Re=50000$ and $Re=5000$ as $\mathrm{ACH} = \left(2 \cdot 1.2 \mathrm{m}^2 \cdot 1.0 \mathrm{m/s} \right) / \left[8 \mathrm{m} \cdot \mathrm{8} \mathrm{m} \cdot \mathrm{3} \mathrm{m} \right] = 45$ 1/h and $\mathrm{ACH} = \left(2 \cdot 1.2 \mathrm{m}^2 \cdot 0.1 \mathrm{m/s} \right) / \left[8 \mathrm{m} \cdot \mathrm{8} \mathrm{m} \cdot \mathrm{3} \mathrm{m} \right] = 4.5$ 1/h, respectively.
The computational ACH values for the various cases can be acquired by imposing a fit of the form $f(t) = C_0 \exp(-\mathrm{ACH}_{fit} \cdot t) + C_1$ to the data presented in Fig.~\ref{fig:fig4_3}. These values, displayed in Tab.~\ref{tab:tab4}, vary between 82\% - 150\% ($Re=50000$) and 75\% - 139\% ($Re=5000$) of the theoretical mixing ventilation value. For $Re=50000$, the displayed ACH values are high (35-70) yet the values are consistent with some of the reported values obtained in experimental settings involving natural ventilation~\cite{escombe2007natural,qian2010natural}. For $Re=5000$, the ACH values (3.4-6.2) correspond to typical values observed in mechanical ventilation setups. For both Reynolds numbers, the ventilation generated by the ceiling vents (1RV) outperforms the various cross-draught setups (3RW, 1RW) and among the two cross-draught simulations, the space with room dividers exhibits enhanced ventilation performance. This is due to the improved mixing of the air masses via a combination of jet impingement, turbulence and flow re-circulation within the back room, which were already discussed in conjunction with Figs.~\ref{fig:fig4_31} and~\ref{fig:fig4_23}.
This suggests that the net impact of the solid obstacles on the ACH value is more ambiguous than one might anticipate as it may depend on the exact details of the airflow and airflow-obstacle interactions. However, the present numerical findings at $Re=5000$ and $Re=50000$ based on full 3D numerical data imply that the theoretical ACH value may be highly inaccurate and off-set by a factor of 0.75-1.51.
\begin{table}[t]
    \centering
    \begin{tabular}{c|c|c|c} 
         Simulation & ACH & ACH$_{\mathrm{min}}$- ACH$_{\mathrm{max}}$ & $\gamma_{\mathrm{min}}$ ... $\gamma_{\mathrm{max}}$ \\ \hline
         DNSLABIB - 1RW & 36.9/3.48 & 35.8 -- 37.9 / 3.38 -- 3.59 & 0.80 -- 0.84 / 0.75 -- 0.80\\
         DNSLABIB - 1RV & 67.6/6.21 & 67.4 -- 67.8 / 6.17 -- 6.24 & 1.50 -- 1.51 / 1.37 -- 1.39\\
         DNSLABIB - 3RW & 48.8/5.00 & 48.2 -- 49.4 / 4.94 -- 5.06 & 1.07 -- 1.10 / 1.10 -- 1.12 \\
         OpenFOAM - 3RW & 50.5/4.53 & 50.0 -- 51.0 / 4.51 -- 4.54 & 1.11 -- 1.13 / 1.00 -- 1.01 \\
         Theoretical value & \underline{45.0} / \underline{4.5} & -- & -- \\
    \end{tabular}
    \caption{The ACH values estimated for $Re=5000$/$50000$ simulations via curve fitting $f(t) = C_0 \exp(-\mathrm{ACH}\cdot t) + C_1$ to each $\langle \mathrm{CO}_2 \rangle$ displayed in Fig.~\ref{fig:fig4_3}. The last two columns detail the range of ACH values based on the error estimates (see Fig. ~\ref{fig:fig4_3}) and their values normalized by the theoretical ACH value.}
    \label{tab:tab4}
\end{table}

\begin{figure}[t]
\centering
 \includegraphics[width=0.88\textwidth]{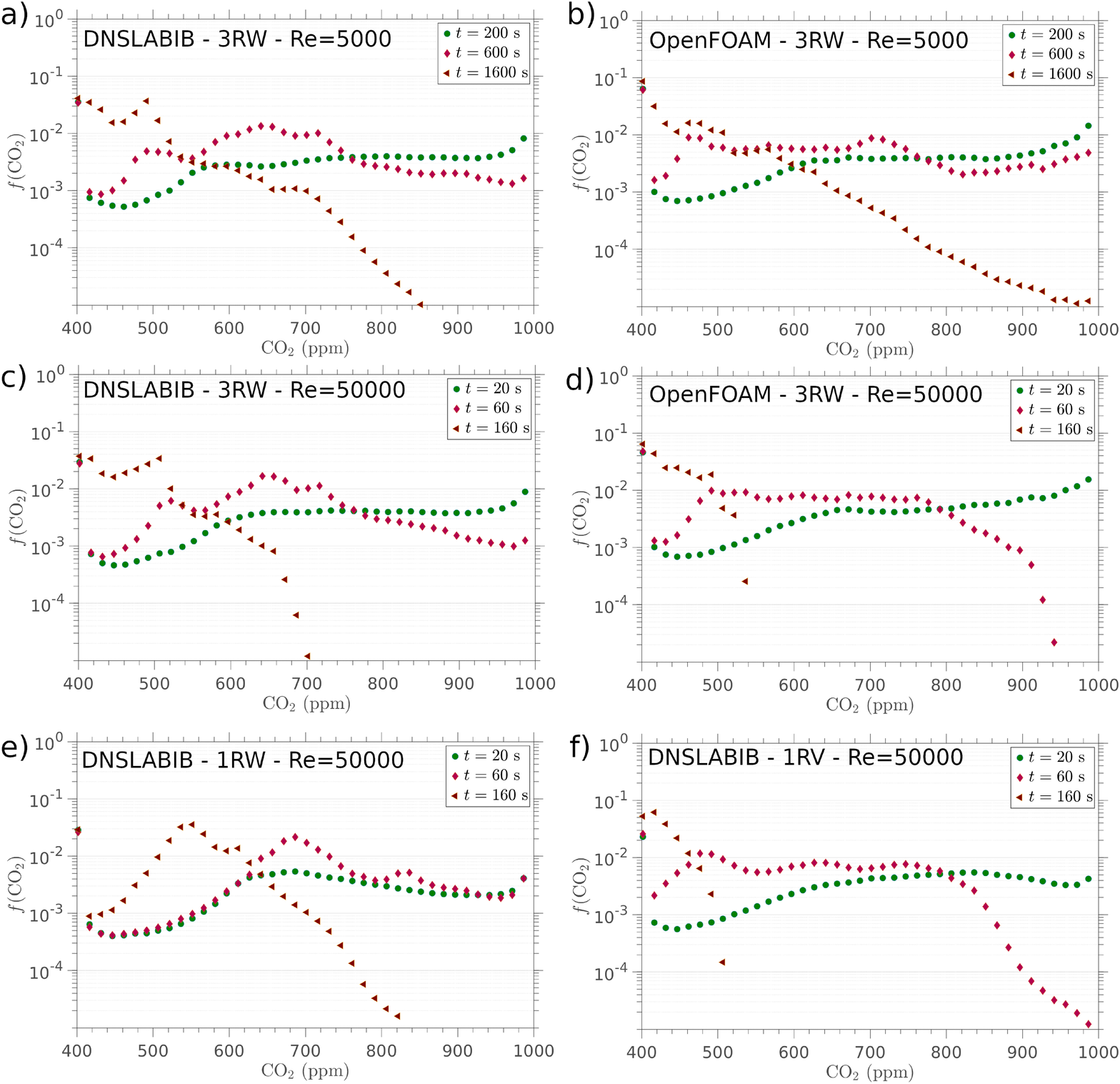}
 \caption{The distribution function $f (\mathrm{CO}_2)$ of the carbon dioxide content at $t=200$ s, $t=600$ s, $t=1600$ s for the 3RW case at $Re=5000$ (low ventilation rate) in DNSLABIB (a) and OpenFOAM (b). Additionally, the distribution is displayed at $t=20$ s, $t=60$ s, $t=160$ for these cases at $Re=50000$ (high ventilation rate) in c) and d). Finally, the profiles are also plotted for the DNSLAB simulations of the 1RW case (e) and 1RV case (f).}
 \label{fig:fig3_2}
\end{figure}
In order to explore the fine structures in Fig.~\ref{fig:fig4_31} and the CO$_2$ profiles they entail, the distribution of $\mathrm{CO}_2$ content in these cases is examined in Fig.~\ref{fig:fig3_2}, where the normalized distribution function $f(\mathrm{CO}_2)$ is plotted for different instances of time, where $t=0$ s denotes the start of the simulation. The respective DNSLABIB and OpenFOAM results for 3RW are displayed in a) and b) ($Re=5000$) and c) and d) ($Re=50000$) while the profiles from the 1RW and 1RV simulations obtained with DNSLABIB are plotted in c) and d) ($Re=50000$). Initially, the probability distribution peaks at CO$_2 = 1000$ ppm. As the simulation and the state of ventilation progresses, the distribution veers towards the lower end of the $\mathrm{CO}_2$ spectrum as anticipated. Furthermore, the distribution functions clearly imply non-homogeneous mixing of the air, perceived as flat and uniform distributions, supporting the observations made earlier.

Therefore, based on the present numerical findings, we find that the airflow significantly affects the mixing patterns and dilution of the CO$_2$ concentration in the configurations considered here. The routinely employed definition of ACH is limited to conditions of perfect and extremely rapid mixing, rarely encountered in realistic indoor ventilation setups. In reality, as exemplified by the results in Fig.~\ref{fig:fig4_3}, the standard deviation in CO$_2$ concentration levels can be in the order of $10-20$\% of the mean value or higher, also pointing to non-homogeneous mixing of the air masses. However, the numerical results indicate that the main difference between the present empty rooms stems from the flow geometry and air supply configuration while the local variation of ACH can be relatively important as well. In the presence of large pieces of furniture ({\it{e.g.}} book shelves) or room dividers, the local variation of ACH may become more prominent which could be considered more in the future.   

\subsection{Infection risk}
Having extracted the ACH values from each simulation case by numerical fitting in the previous discussion, we next proceed to the more practical implications of these results in terms of the infection risk. Therefore, a framework for relating the ACH and infection risk is required. During the COVID-19 pandemic, the classical Wells-Riley equation has been utilized extensively for infection risk assessment ~\cite{auvinen2022high,dai2020association,sun2020efficacy,foster2021estimating,park2021natural,peng2021exhaled,wang2022coupled}. According to the Wells-Riley model, the infection probability ($P_{inf}$) can be calculated as follows
\begin{equation}
    P_{inf} = 1-\exp(\frac{-I q p t}{{\dot V}}), 
\end{equation}
 where $I$ is the number of infectious people in the modeled setting, $q$ is the rate of generation of the infectious units termed "quanta", $p$ is the respiratory rate of a person, $t$ is the exposure time and ${\dot V}$ is the air exchange rate (in units of [m$^3$/h]). This form of the model assumes 1) a steady state situation reached over a longer period of time during which the infectious emit virus to the air, 2) immediate and uniform mixing in the room so that distance from the source is not taken into account, and 3) constant removal of airborne particles by the ventilation. As a remark, under steady state conditions, the argument in the exponential function is simply the inhaled dose which is proportional to the average concentration ([$c_q$]=1/m$^3$) of quanta in the room air which ($I$=1) can be calculated simply as follows \cite{vuorinen2020modelling}
 \begin{equation}
    c_q = \frac{q}{\dot{V}} = \frac{q}{ACH \cdot V}. 
\end{equation}
 
 Here, we address the infection risk in the three indoor settings, assuming that an infectious person has occupied the space for a period of time, saturated the room with exhaled air (high CO$_2$ content) and dispersed infectious quanta to the space which remain infectious. Then, the infected person leaves the room, ventilation is commenced and the infection risk for a person entering the room starts accumulating. We therefore rewrite the Wells-Riley model as follows 
\begin{align}
    P_{inf} &= 1-\exp(-Q(t)) , \\
    Q(t) &= p \int_0^t c_q (t) = C_0 p \int_0^t \exp(-\mathrm{ACH} \cdot t) dt = \frac{C_0 p}{\mathrm{ACH}} \left[1-\exp(-\mathrm{ACH}\cdot t) \right] \label{eq:W-R},
\end{align}
where $Q(t)$ is the effective dose a person has accumulated during time $t$ and $c_q (t)$ is the time-dependent average concentration of quanta in the room air. We assume that $c_q$ is directly proportional to the previously discussed mean CO$_2$ content, {\it{i.e.}} $c_q (t) = C_0 \exp(-\mathrm{ACH} \cdot t)$, where $C_0$ represents the initial, homogeneous concentration of quanta in the room saturated with stale air. This assumption considers only the small aerosols which remain airborne for very long periods of time (see next section). 

\begin{figure}[h!]
\centering
 \includegraphics[width=0.96\textwidth]{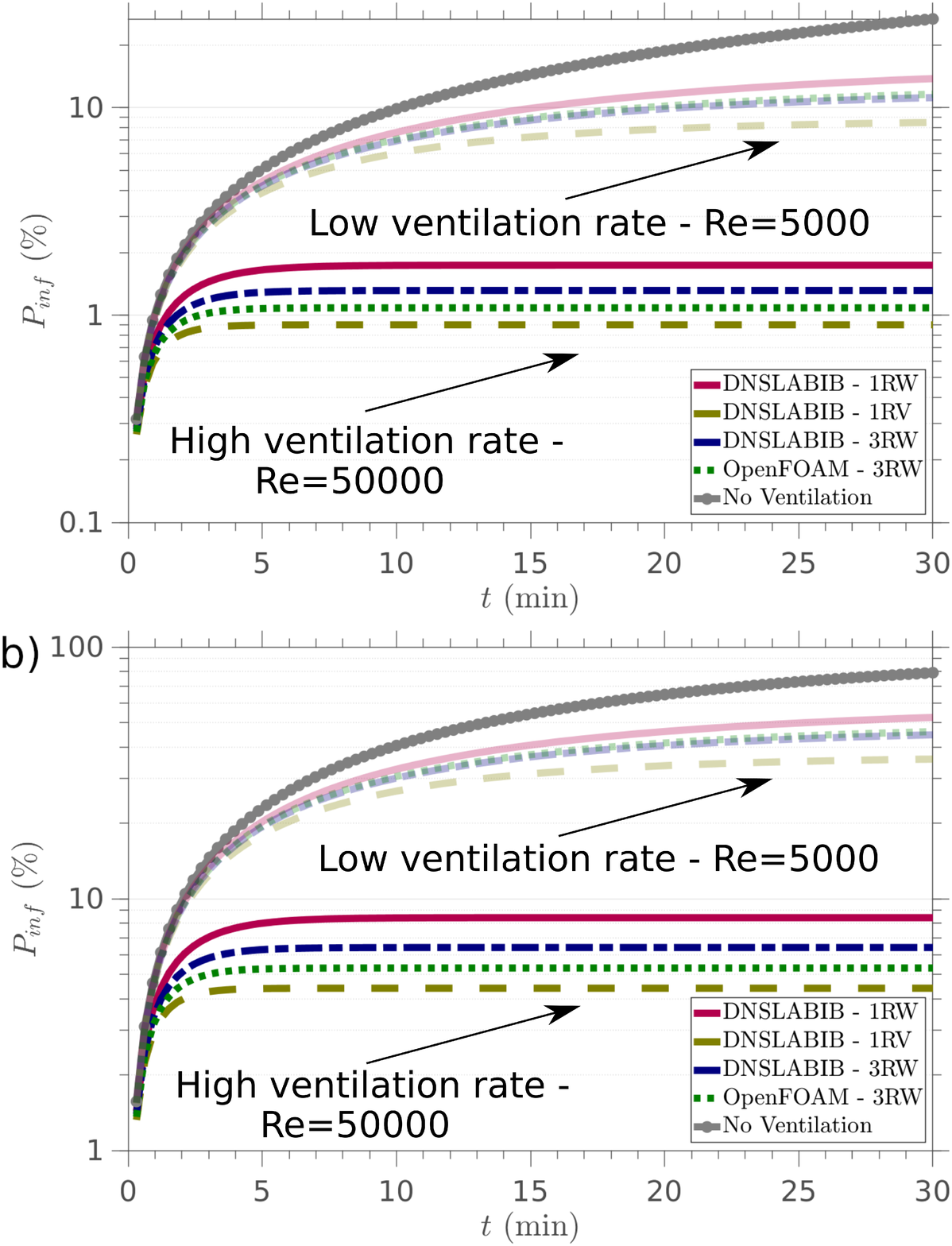}
 \caption{The average infection risk accumulates as a function of time. Here, a scenario is illustrated where an infectious person has released virus to the air and a susceptible person arrives to the room at time $t=0$ s. Initial conditions (a) $C_0 = 100$ $\mathrm{quanta} / V_{room}$ and (b) $C_0 = 500$ $\mathrm{quanta} / V_{room}$. All configurations are highly effective in reducing the risk of infection at higher ventilation rates. At lower ventilation rates (transparent lines), the risk reduction is also significant.}
 \label{fig:fig4_4}
\end{figure}

Fig.~\ref{fig:fig4_4} a) presents the infection risk based on Eq.~\eqref{eq:W-R} with the initial values of (a) 100 quanta  and (b) 500 quanta homogeneously spread in the room volume, $C_0 = \{100, 500\} / (8 \cdot 8 \cdot 3) \approx \{0.52, 2.60\}$ quanta/m$^3$. In the COVID-19 context~\cite{buonanno2020estimation,buonanno2020quantitative} such values could be representative to a person performing activities in an indoor setting over a 1 hour period, releasing pathogens either at a moderate (medium vocal activities, such as talking) or very high rate (singing), respectively. The present demonstrative cases are displayed for $Re=50000$ and $Re=5000$ plotted with solid and transparent curves respectively. Furthermore, as the respiration rate of a person, the value of $p = 1.2 \mathrm{m}^3 / \mathrm{h}$ is applied herein~\cite{vuorinen2020modelling}. The key observation from panel a) is that a high enough ventilation rate reduces the average infection risk extremely efficiently. 

The average infection risk approaches $\approx$ 24\% probability level if ventilation is switched off for the considered time window of 30 minutes compared to the probability of 0.9-1.9 \% calculated for the well-ventilated cases. Even in the worst-performing ventilation setup (1RW), the infection risk reduces by a factor of 10. Similar reductions are also apparent in the results presented in panel b), where the absolute reduction in the infection risk is even greater ($\approx$ 80 \%). For the lower ventilation rate  ($Re=5000$),  the risk is typically reduced by a factor of 2 for each setup. For instance, in a), the ventilation setups at low inflow rate reduce the infection risk from 24 \% to  9-13 \% and in b), from 80 \% to around 37-53 \%. Yet, this can be considered to be a significant gain. As demonstrated herein, window ventilation and enhancement of mechanical ventilation could be considered to be powerful complementary tools in keeping the society open and reducing infection risks in public places such as schools, choir practices, shops and bars. As an additional note, HEPA filters deliver clean air at a certain volumetric flow rate and in analogy with ACH definition, an effective air change eACH (volume of filtered air / room volume) can be defined. HEPA filters offer an energy efficient way in reducing indoor virus concentration to increase the effective ACH value ACH$^{*}$=ACH + eACH and correspondingly lower the virus concentration indoors. In the future, DNSLABIB could be used to model air filtration devices positioned at different indoor locations via the volumetric source terms. 

\subsection{Dispersion of airborne particles}
In the previous section, we discussed the reduced infection risk associated with various ventilation setups. The estimates considered only the fraction of the airborne particles which are small enough so that they can be directly correlated with the indoor CO$_2$ concentration. But what size particles are small enough to follow the airflow? During the COVID-19 pandemic there has been a major scientific debate on the distinction between an aerosol and a droplet. In the aerosol community it is well understood that particles up to sizes of 50-100 $\mu m$ are able to easily remain airborne over extended times and distances because they evaporate quickly~\cite{vuorinen2020modelling}. However, until COVID-19, the medical literature, including WHO in their early guidance in 3/2020, adhered to an erroneous 5 $\mu$m cutoff. Such an unfortunate misconception biased the early attention towards surface transmission which was a major error corrected later on in the pandemic when COVID-19 was noted to be airborne by WHO~\cite{who2021}.  Presently, there is a broad scientific consensus on the airborne route as a major driver for the ongoing pandemic~\cite{tang2020aerosol,greenhalgh2021ten}. Next, we discuss how solid particles between 1-100 $\mu$m travel in the air using DNSLABIB.

\begin{figure}[h!]
\centering
 \includegraphics[width=0.98\textwidth]{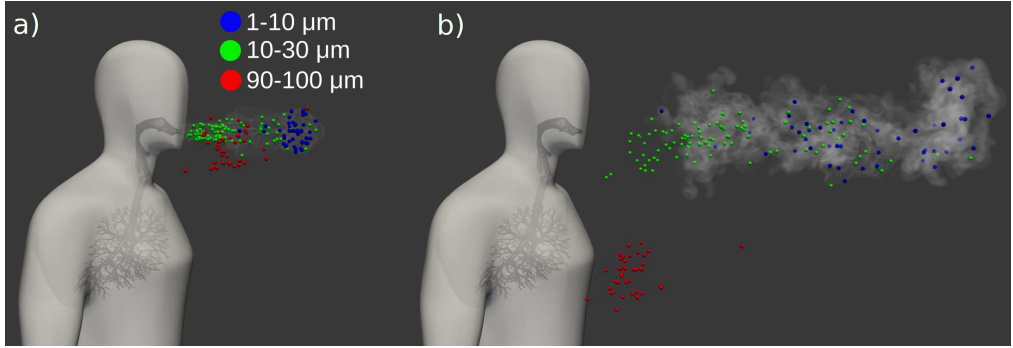}
 \caption{Here, an exhalation is modeled as a particle laden jet. In this particular example, solid and non-evaporating particles of size 30 $\mu$m are noted to remain airborne and easily travel horizontally to reach the airways of another person. For practical mucus droplets, evaporation shifts the critical size to a much larger value, up to 100 $\mu$m \cite{vuorinen2020modelling, tang2020aerosol}.}
 \label{fig:fig5}
\end{figure}

Fig.~\ref{fig:fig5} presents two time snapshots of an exhaling person 0.2 seconds (a) and 0.8 seconds (b) after start of exhalation. The air pulse is modelled as a particle laden jet with a diameter of $D = 3$ cm. The Reynolds number of the exhaled jet is $Re=7000$. The gray cloud represents the passive scalar field generated in this expulsion of air posing high CO$_2$ concentration while the red/green/blue droplets present large (90-100 $\mu$m), medium (10-30 $\mu$m) and small (1-10 $\mu$m) particles emitted from the airways, respectively. These size classes are chosen in order to comply with the clinical evidence from human expired aerosol measurements~\cite{johnson2011modality}. The distribution of the particle sizes in our simulation is 20\% (90-100 $\mu$m), 53.3\% (10-30 $\mu$m) and 26.7\% (1-10 $\mu$m) of the total particle count (200), respectively. Here, particle evaporation is neglected so that the observed picture corresponds to a conservative situation where particles sink significantly faster than they would sink in reality e.g. at a RH=30-40\% indoor relative humidity \cite{vuorinen2020modelling}. 

Notably, the smaller particles up to 30 $\mu$m remain airborne and they are transported easily through the air by the spreading air jet for considerable distances, the trajectories having a clear correlation to the exhaled CO$_2$ plume. This is consistent with our previous findings~\cite{vuorinen2020modelling} where particles of size 20 $\mu$m were shown to be transported over shelves of a supermarket between two aisles. However, the particle behavior also simply follows from the the particle sedimentation time in still air with $\tau_s$ being 36 min, 9 min and 1 min for particle sizes of 5 $\mu$m, 10 $\mu$m and 30 $\mu$m, respectively.

\begin{figure}[h!]
\centering
 \includegraphics[width=0.92\textwidth]{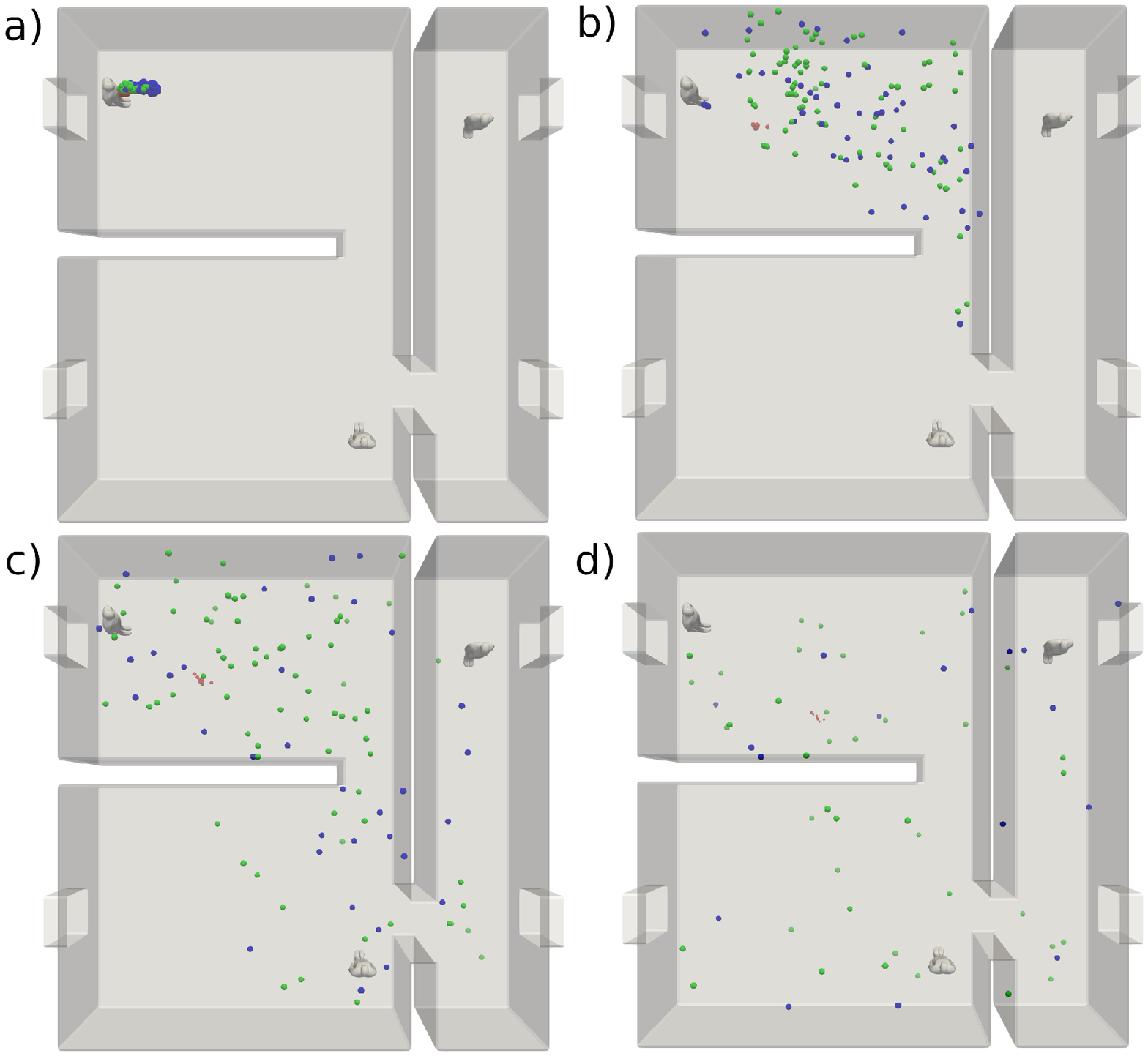}
 \caption{A simulation of a room with the person emitting infectious aerosols located in the top left corner (3RW case at $Re=50000$, high ventilation rate). The a), b), c) and d) panels correspond to the simulation times $t=0$ s, $t=20$ s, $t=40$ s and $t=100$ s, respectively after the onset of aerosol emission by the person.}
 \label{fig:fig6}
\end{figure}
Finally, Fig.~\ref{fig:fig6} illustrates the aerosol emitting person placed in a room with a cross-draught and room dividers (3RW). The person emitting the infectious aerosols is located in the left upper corner of the room at a) $t=0$ s, b) $t=20$ s, c) $t=40$ s and d) $t=100$ s  respectively. The simulations suggest that the smallest aerosols ($\leq$ 30 $\mu$m) can rapidly travel significant distances following the airflow, bypassing the solid walls.

The results displayed here also suggest revisiting social distancing guidelines and protection measures in combating infectious respiratory disease. Within one minute, the two individuals separated from the initial aerosol source by two large walls have been exposed to the smallest aerosols.
Since the aerosols travel from room to room and further to the corridor, this example clearly supports the notion that in general, room dividers, plexi-glasses or visors do not provide sufficient protection from aerosols.   
\section{Conclusions}
In the present work, we endeavored to create a CFD tool for scale-resolved simulations with reduced computational effort. Therefore, the DNSLABIB software was programmed in the MATLAB language and implemented on a GPU. Extending the capabilities of MATLAB to the realm of CFD, DNSLABIB provides the end-user with prior CFD experience a starting point for further exploration.

DNSLABIB implements solid obstacles in a simplified manner via the Immersed Boundary (IB) method and the simulations are executed on a GPU. DNSLABIB was validated in two canonical reference cases, the pressure-driven channel flow and a channel flow past a bluff body. The code was utilized to study three separate indoor ventilation configurations using scale-resolving simulations. In one of these cases, a comparison against a respective OpenFOAM simulation was performed as well. A superior performance of DNSLABIB over the corresponding OpenFOAM implementation was discovered, the speed-up factor being in the order of 3 while avoiding usage of a supercomputer which is considered as a major step advocating the usage of GPUs for indoor airflow assessment. 

The ventilation characteristics in the three cases were studied by monitoring the CO$_2$ concentration as it is routinely adopted as a measured proxy for airborne transmission risk. The CO$_2$ distributions and mean value over time revealed highly inhomogeneous mixing of air, contrasting the common assumption of ideally mixed air employed to determine the ACH value. The actual ACH for each case was determined and significant deviations from the theoretical value were noted. Collectively, the results indicated the presence of strong local variations in the CO$_2$ concentration indoors. This further emphasizes the need to better understand the full 3D airflow features indoors using scale-resolved simulations to ensure high air quality both locally and on average. 

The extracted ACH parameters were further applied in the Wells-Riley model to assess the infection risk associated with each ventilation setup in the context of COVID-19. The provided examples indicate significantly reduced infection risk if windows are opened immediately after entering a room with stale, virus-contaminated air. However, it should be highlighted that the actual benefits of ventilation emerge only if the  inhalation dose $Q$ remains small enough which can be achieved by 1) minimizing exposure time and  2) minimizing airborne virus concentration via enhanced ventilation and/or air filtration or masks. Even in the most conservative scenarios with the lowest ventilation rates (ACH=3.4-6.3), we provided numerical examples on cases where the infection risk was reduced by a factor of 2, which is very significant {\it{per se}}. For the well ventilated cases (ACH=37-67), over a 10-fold decrease in the infection risk was observed.

Finally, the transportation of respiratory particles in an exhaled jet was investigated. The studied condition resembles a case with a relative humidity of RH=100\% where particle size reduction due to evaporation does not occur~\cite{vuorinen2020modelling}. In such a conservative setup, solid particles up to 30 $\mu$m in size were witnessed to remain airborne for prolonged periods of time and they were therefore shown to affect observers in the considered premises over extended distances as well. It is clear that the traditional 5 $\mu$m cutoff is simply erroneous. 
However, the numerical example in question indicates that the large particles with sedimentation time on the order of $\sim$1-10 seconds will exit the exhaled air jet region quickly after which they will settle on the floor and surfaces. For large particles (here:  90-100$\mu$m), it is clear that the direction and strength of the exhaled and the ambient airflow will play a major role in how well those particles may reach other people's airways near-by before settling down. The medium particles (here: 10-30 $\mu$m) are clearly able to stay airborne for a long duration of time and travel over extended distances. The smallest particles (here: $<$10 $\mu$m) are able to remain airborne for very long periods of time. The inhalation of small and medium aerosols is the most likely route of virus transmission as they can be unconditionally inhaled over short and long distances and their viral content~\cite{wang2021airborne} as well as their number-concentration is abundant~\cite{johnson2011modality}.  

The complexity underlying the ventilation cases studied here also suggests numerous venues for future research. A logical continuation could be the development of a wall model to more accurately consider boundary layers at high shear surfaces. However, as noted here in particular for $Re=5000$, this aspect may not be completely critical for low-speed indoor airflows if most of the near-wall $y+$ values remain small enough. Additionally, since indoor airflow can be considerably affected by buoyancy effects, such as heat generated by the occupants, adding an appropriate coupling (e.g. Boussinesq approximation) between the heat sources (sinks) and the momentum equation would be reasonable as well.   

\section*{Code availability}
\noindent The DNSLABIB package is freely available at https://github.com/Aalto-CFD/DNSLABIB.

\section*{Acknowledgements}
\noindent We thank the Academy of Finland for their financial support (grant No. 335516) and the Aalto Science-IT project for the high-performance computational resources.

%%%REFERENCES%%%
\bibliographystyle{elsarticle-num}
\bibliography{master}

\begin{thebibliography}{10}
\expandafter\ifx\csname url\endcsname\relax
  \def\url#1{\texttt{#1}}\fi
\expandafter\ifx\csname urlprefix\endcsname\relax\def\urlprefix{URL }\fi
\expandafter\ifx\csname href\endcsname\relax
  \def\href#1#2{#2} \def\path#1{#1}\fi

\bibitem{wang2021airborne}
C.~C. Wang, K.~A. Prather, J.~Sznitman, J.~L. Jimenez, S.~S. Lakdawala,
  Z.~Tufekci, L.~C. Marr, Airborne transmission of respiratory viruses, Science
  373~(6558) (2021) eabd9149.

\bibitem{tellier2022covid}
R.~Tellier, {COVID-19}: the case for aerosol transmission, Interface Focus
  12~(2) (2022) 20210072.

\bibitem{auvinen2022high}
M.~Auvinen, J.~Kuula, T.~Gr{\"o}nholm, M.~S{\"u}hring, A.~Hellsten,
  High-resolution large-eddy simulation of indoor turbulence and its effect on
  airborne transmission of respiratory pathogens—model validation and
  infection probability analysis, Physics of Fluids 34~(1) (2022) 015124.

\bibitem{anderson2020consideration}
E.~L. Anderson, P.~Turnham, J.~R. Griffin, C.~C. Clarke, Consideration of the
  aerosol transmission for {COVID-19} and public health, Risk Analysis 40~(5)
  (2020) 902--907.

\bibitem{tang2020aerosol}
S.~Tang, Y.~Mao, R.~M. Jones, Q.~Tan, J.~S. Ji, N.~Li, J.~Shen, Y.~Lv, L.~Pan,
  P.~Ding, et~al., Aerosol transmission of {SARS-CoV-2}? evidence, prevention
  and control, Environment international 144 (2020) 106039.

\bibitem{jayaweera2020transmission}
M.~Jayaweera, H.~Perera, B.~Gunawardana, J.~Manatunge, Transmission of
  {COVID-19} virus by droplets and aerosols: A critical review on the
  unresolved dichotomy, Environmental research (2020) 109819.

\bibitem{mittal2020flow}
R.~Mittal, R.~Ni, J.-H. Seo, The flow physics of {COVID-19}, Journal of fluid
  Mechanics 894 (2020).

\bibitem{fears2020persistence}
A.~C. Fears, W.~B. Klimstra, P.~Duprex, A.~Hartman, S.~C. Weaver, K.~S. Plante,
  D.~Mirchandani, J.~A. Plante, P.~V. Aguilar, D.~Fern{\'a}ndez, et~al.,
  Persistence of severe acute respiratory syndrome coronavirus 2 in aerosol
  suspensions, Emerging infectious diseases 26~(9) (2020) 2168.

\bibitem{van2020aerosol}
N.~Van~Doremalen, T.~Bushmaker, D.~H. Morris, M.~G. Holbrook, A.~Gamble, B.~N.
  Williamson, A.~Tamin, J.~L. Harcourt, N.~J. Thornburg, S.~I. Gerber, et~al.,
  Aerosol and surface stability of {SARS-CoV-2} as compared with {SARS-CoV-1},
  New England journal of medicine 382~(16) (2020) 1564--1567.

\bibitem{zhang2020identifying}
R.~Zhang, Y.~Li, A.~L. Zhang, Y.~Wang, M.~J. Molina, Identifying airborne
  transmission as the dominant route for the spread of {COVID-19}, Proceedings
  of the National Academy of Sciences 117~(26) (2020) 14857--14863.

\bibitem{wilson2020airborne}
N.~Wilson, A.~Norton, F.~Young, D.~Collins, Airborne transmission of severe
  acute respiratory syndrome coronavirus-2 to healthcare workers: a narrative
  review, Anaesthesia 75~(8) (2020) 1086--1095.

\bibitem{godri2020covid}
K.~J. Godri~Pollitt, J.~Peccia, A.~I. Ko, N.~Kaminski, C.~S. Dela~Cruz, D.~W.
  Nebert, J.~K. Reichardt, D.~C. Thompson, V.~Vasiliou, {COVID-19}
  vulnerability: the potential impact of genetic susceptibility and airborne
  transmission, Human genomics 14 (2020) 1--7.

\bibitem{li2021probable}
Y.~Li, H.~Qian, J.~Hang, X.~Chen, P.~Cheng, H.~Ling, S.~Wang, P.~Liang, J.~Li,
  S.~Xiao, et~al., Probable airborne transmission of {SARS-CoV-2} in a poorly
  ventilated restaurant, Building and Environment (2021) 107788.

\bibitem{henriques2021modelling}
A.~Henriques, N.~Mounet, L.~Aleixo, P.~Elson, J.~Devine, G.~Azzopardi,
  M.~Andreini, M.~Rognlien, N.~Tarocco, J.~Tang, Modelling airborne
  transmission of {SARS-CoV-2} using {CARA}: risk assessment for enclosed
  spaces, Interface Focus 12~(2) (2022) 20210076.

\bibitem{eames2022spread}
I.~Eames, J.-B. Fl{\'o}r, Spread of infectious agents through the air in
  complex spaces, Interface Focus 12~(2) (2022) 20210080.

\bibitem{ascione2021design}
F.~Ascione, R.~F. De~Masi, M.~Mastellone, G.~P. Vanoli, The design of safe
  classrooms of educational buildings for facing contagions and transmission of
  diseases: A novel approach combining audits, calibrated energy models,
  building performance ({BPS}) and computational fluid dynamic ({CFD})
  simulations, Energy and Buildings 230 (2021) 110533.

\bibitem{zhang2021simulation}
F.~Zhang, Y.~Ryu, Simulation study on indoor air distribution and indoor
  humidity distribution of three ventilation patterns using computational fluid
  dynamics, Sustainability 13~(7) (2021) 3630.

\bibitem{abuhegazy2020numerical}
M.~Abuhegazy, K.~Talaat, O.~Anderoglu, S.~V. Poroseva, Numerical investigation
  of aerosol transport in a classroom with relevance to {COVID-19}, Physics of
  Fluids 32~(10) (2020) 103311.

\bibitem{borro2021role}
L.~Borro, L.~Mazzei, M.~Raponi, P.~Piscitelli, A.~Miani, A.~Secinaro, The role
  of air conditioning in the diffusion of {S}ars-{C}o{V}-2 in indoor
  environments: A first computational fluid dynamic model, based on
  investigations performed at the vatican state children's hospital,
  Environmental Research 193 (2021) 110343.

\bibitem{liu2020modeling}
W.~Liu, T.~van Hooff, Y.~An, S.~Hu, C.~Chen, Modeling transient particle
  transport in transient indoor airflow by fast fluid dynamics with the markov
  chain method, Building and Environment 186 (2020) 107323.

\bibitem{buchan2020predicting}
A.~G. Buchan, L.~Yang, K.~D. Atkinson, Predicting airborne coronavirus
  inactivation by far-{UVC} in populated rooms using a high-fidelity coupled
  radiation-{CFD} model, Scientific reports 10~(1) (2020) 1--7.

\bibitem{vuorinen2020modelling}
V.~Vuorinen, M.~Aarnio, M.~Alava, V.~Alopaeus, N.~Atanasova, M.~Auvinen,
  N.~Balasubramanian, H.~Bordbar, P.~Er{\"a}st{\"o}, R.~Grande, et~al.,
  Modelling aerosol transport and virus exposure with numerical simulations in
  relation to {SARS-CoV-2} transmission by inhalation indoors, Safety Science
  130 (2020) 104866.

\bibitem{ren2021numerical}
J.~Ren, Y.~Wang, Q.~Liu, Y.~Liu, Numerical study of three ventilation
  strategies in a prefabricated {COVID-19} inpatient ward, Building and
  Environment 188 (2021) 107467.

\bibitem{dbouk2021airborne}
T.~Dbouk, D.~Drikakis, On airborne virus transmission in elevators and confined
  spaces, Physics of Fluids 33~(1) (2021) 011905.

\bibitem{ho2021modeling}
C.~K. Ho, Modeling airborne pathogen transport and transmission risks of
  {SARS-CoV-2}, Applied mathematical modelling 95 (2021) 297--319.

\bibitem{ho2021modelling}
C.~K. Ho, Modelling airborne transmission and ventilation impacts of a
  {COVID-19} outbreak in a restaurant in guangzhou, china, International
  Journal of Computational Fluid Dynamics (2021) 1--19.

\bibitem{li2020investigating}
H.~Li, K.~Zhong, Z.~J. Zhai, Investigating the influences of ventilation on the
  fate of particles generated by patient and medical staff in operating room,
  Building and Environment 180 (2020) 107038.

\bibitem{khosronejad2020fluid}
A.~Khosronejad, C.~Santoni, K.~Flora, Z.~Zhang, S.~Kang, S.~Payabvash,
  F.~Sotiropoulos, Fluid dynamics simulations show that facial masks can
  suppress the spread of {COVID-19} in indoor environments, AIP Advances
  10~(12) (2020) 125109.

\bibitem{who2021}
W.~H.~O. ({WHO}),
  \href{https://www.who.int/news-room/questions-and-answers/item/coronavirus-disease-covid-19-how-is-it-transmitted}{Coronavirus
  disease ({COVID-19}): How is it transmitted?}
\newline\urlprefix\url{https://www.who.int/news-room/questions-and-answers/item/coronavirus-disease-covid-19-how-is-it-transmitted}

\bibitem{nielsen2015fifty}
P.~V. Nielsen, Fifty years of {CFD} for room air distribution, Building and
  Environment 91 (2015) 78--90.

\bibitem{blocken2018over}
B.~Blocken, {LES} over {RANS} in building simulation for outdoor and indoor
  applications: a foregone conclusion?, in: Building Simulation, Vol.~11,
  Springer, 2018, pp. 821--870.

\bibitem{pratx2011gpu}
G.~Pratx, L.~Xing, {GPU} computing in medical physics: A review, Medical
  physics 38~(5) (2011) 2685--2697.

\bibitem{niemeyer2014recent}
K.~E. Niemeyer, C.-J. Sung, Recent progress and challenges in exploiting
  graphics processors in computational fluid dynamics, The Journal of
  Supercomputing 67~(2) (2014) 528--564.

\bibitem{liu2004real}
Y.~Liu, X.~Liu, E.~Wu, Real-time {3D} fluid simulation on {GPU} with complex
  obstacles, in: 12th Pacific Conference on Computer Graphics and Applications,
  2004. PG 2004. Proceedings., IEEE, 2004, pp. 247--256.

\bibitem{scheidegger2005practical}
C.~E. Scheidegger, J.~L. Comba, R.~D. Da~Cunha, Practical {CFD} simulations on
  programmable graphics hardware using {SMAC}, in: Computer Graphics Forum,
  Vol.~24, Wiley Online Library, 2005, pp. 715--728.

\bibitem{shinn2009implementation}
A.~F. Shinn, S.~P. Vanka, Implementation of a semi-implicit pressure-based
  multigrid fluid flow algorithm on a graphics processing unit, in: ASME
  International Mechanical Engineering Congress and Exposition, Vol. 43864,
  2009, pp. 125--133.

\bibitem{thibault2009cuda}
J.~Thibault, I.~Senocak, {CUDA} implementation of a {N}avier-{S}tokes solver on
  multi-{GPU} desktop platforms for incompressible flows, in: 47th AIAA
  aerospace sciences meeting including the new horizons forum and aerospace
  exposition, 2009, p. 758.

\bibitem{brandvik2011turbo}
T.~Brandvik, G.~Pullan, An accelerated {3D} {N}avier–{S}tokes solver for
  flows in turbomachines, ASME. J. Turbomach. 133~(2) (2010) 021025.

\bibitem{griebel2010multi}
M.~Griebel, P.~Zaspel, A multi-{GPU} accelerated solver for the
  three-dimensional two-phase incompressible {N}avier-{S}tokes equations,
  Computer Science-Research and Development 25~(1) (2010) 65--73.

\bibitem{zaspel2013solving}
P.~Zaspel, M.~Griebel, Solving incompressible two-phase flows on multi-{GPU}
  clusters, Computers \& Fluids 80 (2013) 356--364.

\bibitem{kelly2014numerical}
J.~M. Kelly, E.~A. Divo, A.~J. Kassab, Numerical solution of the two-phase
  incompressible {N}avier--{S}tokes equations using a {GPU}-accelerated
  meshless method, Engineering Analysis with Boundary Elements 40 (2014)
  36--49.

\bibitem{shinn2010direct}
A.~Shinn, S.~Vanka, W.-m. Hwu, Direct numerical simulation of turbulent flow in
  a square duct using a graphics processing unit ({GPU}), in: 40th Fluid
  Dynamics Conference and Exhibit, 2010, p. 5029.

\bibitem{salvadore2013gpu}
F.~Salvadore, M.~Bernardini, M.~Botti, {GPU} accelerated flow solver for direct
  numerical simulation of turbulent flows, Journal of Computational Physics 235
  (2013) 129--142.

\bibitem{khajeh2013direct}
A.~Khajeh-Saeed, J.~B. Perot, Direct numerical simulation of turbulence using
  {GPU} accelerated supercomputers, Journal of Computational Physics 235 (2013)
  241--257.

\bibitem{shi2012accelerating}
Y.~Shi, W.~H. Green, H.-W. Wong, O.~O. Oluwole, Accelerating multi-dimensional
  combustion simulations using {GPU} and hybrid explicit/implicit {ODE}
  integration, Combustion and Flame 159~(7) (2012) 2388--2397.

\bibitem{spafford2009accelerating}
K.~Spafford, J.~Meredith, J.~Vetter, J.~Chen, R.~Grout, R.~Sankaran,
  Accelerating {S3D}: a {GPGPU} case study, in: European Conference on Parallel
  Processing, Springer, 2009, pp. 122--131.

\bibitem{witherden2014pyfr}
F.~D. Witherden, A.~M. Farrington, P.~E. Vincent, {PyFR}: An open source
  framework for solving advection--diffusion type problems on streaming
  architectures using the flux reconstruction approach, Computer Physics
  Communications 185~(11) (2014) 3028--3040.

\bibitem{vermeire2017utility}
B.~C. Vermeire, F.~D. Witherden, P.~E. Vincent, On the utility of {GPU}
  accelerated high-order methods for unsteady flow simulations: A comparison
  with industry-standard tools, Journal of Computational Physics 334 (2017)
  497--521.

\bibitem{loppi2018high}
N.~A. Loppi, F.~D. Witherden, A.~Jameson, P.~E. Vincent, A high-order
  cross-platform incompressible {N}avier--{S}tokes solver via artificial
  compressibility with application to a turbulent jet, Computer Physics
  Communications 233 (2018) 193--205.

\bibitem{vermeire2019optimal}
B.~C. Vermeire, N.~A. Loppi, P.~E. Vincent, Optimal {R}unge--{K}utta schemes
  for pseudo time-stepping with high-order unstructured methods, Journal of
  Computational Physics 383 (2019) 55--71.

\bibitem{loppi2019locally}
N.~A. Loppi, F.~D. Witherden, A.~Jameson, P.~E. Vincent, Locally adaptive
  pseudo-time stepping for high-order flux reconstruction, Journal of
  Computational Physics 399 (2019) 108913.

\bibitem{vuorinen2016dnslab}
V.~Vuorinen, K.~Keskinen, Dnslab: A gateway to turbulent flow simulation in
  matlab, Computer Physics Communications 203 (2016) 278--289.

\bibitem{nickolls2008scalable}
J.~Nickolls, I.~Buck, M.~Garland, K.~Skadron, Scalable parallel programming
  with cuda: Is cuda the parallel programming model that application developers
  have been waiting for?, Queue 6~(2) (2008) 40--53.

\bibitem{peyk2011electromagnetoencephalography}
P.~Peyk, A.~De~Cesarei, M.~Jungh{\"o}fer, Electromagnetoencephalography
  software: overview and integration with other {EEG}/{MEG} toolboxes,
  Computational intelligence and neuroscience 2011 (2011).

\bibitem{goodman2009brian}
D.~F. Goodman, R.~Brette, The brian simulator, Frontiers in neuroscience 3
  (2009) 26.

\bibitem{daowd2011passive}
M.~Daowd, N.~Omar, P.~Van Den~Bossche, J.~Van~Mierlo, Passive and active
  battery balancing comparison based on {MATLAB} simulation, in: 2011 IEEE
  Vehicle Power and Propulsion Conference, IEEE, 2011, pp. 1--7.

\bibitem{mohanty2014matlab}
P.~Mohanty, G.~Bhuvaneswari, R.~Balasubramanian, N.~K. Dhaliwal, {MATLAB} based
  modeling to study the performance of different {MPPT} techniques used for
  solar {PV} system under various operating conditions, Renewable and
  Sustainable Energy Reviews 38 (2014) 581--593.

\bibitem{alegre2017modelling}
S.~Alegre, J.~V. M{\'\i}guez, J.~Carpio, Modelling of electric and
  parallel-hybrid electric vehicle using {M}atlab/{S}imulink environment and
  planning of charging stations through a geographic information system and
  genetic algorithms, Renewable and Sustainable Energy Reviews 74 (2017)
  1020--1027.

\bibitem{gu2005robust}
D.-W. Gu, P.~Petkov, M.~M. Konstantinov, Robust control design with
  {MATLAB}{\textregistered}, Springer Science \& Business Media, 2005.

\bibitem{wang2009model}
L.~Wang, Model predictive control system design and implementation using
  {MATLAB}{\textregistered}, Springer Science \& Business Media, 2009.

\bibitem{proakis2012contemporary}
J.~G. Proakis, M.~Salehi, G.~Bauch, Contemporary communication systems using
  {MATLAB}, Cengage Learning, 2012.

\bibitem{cho2010mimo}
Y.~S. Cho, J.~Kim, W.~Y. Yang, C.~G. Kang, {MIMO}-{OFDM} wireless
  communications with {MATLAB}, John Wiley \& Sons, 2010.

\bibitem{canuto2007spectral}
C.~Canuto, M.~Y. Hussaini, A.~Quarteroni, T.~A. Zang, Spectral methods:
  evolution to complex geometries and applications to fluid dynamics, Springer
  Science \& Business Media, 2007.

\bibitem{vuorinen2014implementation}
V.~Vuorinen, J.-P. Keskinen, C.~Duwig, B.~J. Boersma, On the implementation of
  low-dissipative {R}unge--{K}utta projection methods for time dependent flows
  using {OpenFOAM}{\textregistered}, Computers \& Fluids 93 (2014) 153--163.

\bibitem{johansen2005dust}
A.~Johansen, H.~Klahr, Dust diffusion in protoplanetary disks by
  magnetorotational turbulence, The Astrophysical Journal 634~(2) (2005) 1353.

\bibitem{peltonen2019large}
P.~Peltonen, K.~Saari, K.~Kukko, V.~Vuorinen, J.~Partanen, Large-eddy
  simulation of local heat transfer in plate and pin fin heat exchangers
  confined in a pipe flow, International Journal of Heat and Mass Transfer 134
  (2019) 641--655.

\bibitem{villanueva2021assessment}
F.~Villanueva, A.~Notario, B.~Caba{\~n}as, P.~Mart{\'\i}n, S.~Salgado, M.~F.
  Gabriel, Assessment of {CO2} and aerosol (pm2. 5, pm10, ufp) concentrations
  during the reopening of schools in the {COVID-19} pandemic: The case of a
  metropolitan area in central-southern spain, Environmental Research 197
  (2021) 111092.

\bibitem{kitamura2021co2}
H.~Kitamura, Y.~Ishigaki, T.~Kuriyama, T.~Moritake, {CO2} concentration
  visualization for {COVID-19} infection prevention in concert halls,
  Environmental and Occupational Health Practice 3~(1) (2021).

\bibitem{poza2021indoor}
I.~Poza-Casado, A.~Llorente-{\'A}lvarez, M.~{\'A}. Padilla-Marcos, Indoor air
  quality in naturally ventilated classrooms. lessons learned from a case study
  in a {COVID-19} scenario, Sustainability 13~(15) (2021) 8446.

\bibitem{chen2021recommendations}
C.-Y. Chen, P.-H. Chen, J.-K. Chen, T.-C. Su, Recommendations for ventilation
  of indoor spaces to reduce {COVID-19} transmission, Journal of the Formosan
  Medical Association 120~(12) (2021) 2055--2060.

\bibitem{schibuola2021high}
L.~Schibuola, C.~Tambani, High energy efficiency ventilation to limit
  {COVID-19} contagion in school environments, Energy and Buildings 240 (2021)
  110882.

\bibitem{bartzanas2007analysis}
T.~Bartzanas, C.~Kittas, A.~Sapounas, C.~Nikita-Martzopoulou, Analysis of
  airflow through experimental rural buildings: Sensitivity to turbulence
  models, Biosystems engineering 97~(2) (2007) 229--239.

\bibitem{escombe2007natural}
A.~R. Escombe, C.~C. Oeser, R.~H. Gilman, M.~Navincopa, E.~Ticona, W.~Pan,
  C.~Mart{\'\i}nez, J.~Chacaltana, R.~Rodr{\'\i}guez, D.~A.~J. Moore, et~al.,
  Natural ventilation for the prevention of airborne contagion, PLoS medicine
  4~(2) (2007) e68.

\bibitem{qian2010natural}
H.~Qian, Y.~Li, W.~Seto, P.~Ching, W.~Ching, H.~Sun, Natural ventilation for
  reducing airborne infection in hospitals, Building and Environment 45~(3)
  (2010) 559--565.

\bibitem{dai2020association}
H.~Dai, B.~Zhao, Association of the infection probability of {COVID-19} with
  ventilation rates in confined spaces, in: Building Simulation, Vol.~13,
  Springer, 2020, pp. 1321--1327.

\bibitem{sun2020efficacy}
C.~Sun, Z.~Zhai, The efficacy of social distance and ventilation effectiveness
  in preventing {COVID-19} transmission, Sustainable cities and society 62
  (2020) 102390.

\bibitem{foster2021estimating}
A.~Foster, M.~Kinzel, Estimating {COVID-19} exposure in a classroom setting: A
  comparison between mathematical and numerical models, Physics of Fluids
  33~(2) (2021) 021904.

\bibitem{park2021natural}
S.~Park, Y.~Choi, D.~Song, E.~K. Kim, Natural ventilation strategy and related
  issues to prevent coronavirus disease 2019 ({COVID-19}) airborne transmission
  in a school building, Science of the Total Environment 789 (2021) 147764.

\bibitem{peng2021exhaled}
Z.~Peng, J.~L. Jimenez, Exhaled {CO2} as a {COVID-19} infection risk proxy for
  different indoor environments and activities, Environmental Science \&
  Technology Letters 8~(5) (2021) 392--397.

\bibitem{wang2022coupled}
Z.~Wang, E.~R. Galea, A.~Grandison, J.~Ewer, F.~Jia, A coupled computational
  fluid dynamics and {W}ells-{R}iley model to predict {COVID-19} infection
  probability for passengers on long-distance trains, Safety science 147 (2022)
  105572.

\bibitem{buonanno2020estimation}
G.~Buonanno, L.~Stabile, L.~Morawska, Estimation of airborne viral emission:
  Quanta emission rate of {SARS-CoV-2} for infection risk assessment,
  Environment international 141 (2020) 105794.

\bibitem{buonanno2020quantitative}
G.~Buonanno, L.~Morawska, L.~Stabile, Quantitative assessment of the risk of
  airborne transmission of {SARS-CoV-2} infection: prospective and
  retrospective applications, Environment international 145 (2020) 106112.

\bibitem{greenhalgh2021ten}
T.~Greenhalgh, J.~L. Jimenez, K.~A. Prather, Z.~Tufekci, D.~Fisman,
  R.~Schooley, Ten scientific reasons in support of airborne transmission of
  {SARS-CoV-2}, The lancet 397~(10285) (2021) 1603--1605.

\bibitem{johnson2011modality}
G.~Johnson, L.~Morawska, Z.~Ristovski, M.~Hargreaves, K.~Mengersen, C.~Y.~H.
  Chao, M.~Wan, Y.~Li, X.~Xie, D.~Katoshevski, et~al., Modality of human
  expired aerosol size distributions, Journal of Aerosol Science 42~(12) (2011)
  839--851.

\end{thebibliography}

% \bibliography{master}

\end{document}